\documentclass[reprint, twocolumn, amsmath, amssymb, aip, jcp, showpacs]{revtex4-1}
\pdfoutput=1
\usepackage[utf8]{inputenc}
\usepackage{graphicx}
\usepackage{color} 

\newcommand{\affwigner}{\affiliation{Institute for Solid State Physics and Optics, Wigner Research Centre for Physics, P.O. Box 49, H-1525 Budapest, Hungary}}

\begin{document}
\title{Ternary eutectic dendrites: Pattern formation and scaling properties}

\author{László Rátkai}
\affwigner

\author{Attila Szállás}
\affwigner

\author{Tamás Pusztai}
\affwigner

\author{Tetsuo Mohri}
\affiliation{Center for Computational Materials Science, Institute for Materials Research, Tohoku University, 2-1-1 Katahira, Aoba-ku, Sendai, 980-8577, Japan}

\author{László Gránásy}
\email{granasy.laszlo@wigner.mta.hu}
\affwigner
\affiliation{Brunel University, Uxbridge, Middlesex, UB8 3PH, United Kingdom}

\pacs{68.70.+w, 81.10.Aj, 81.30.Fb}

\begin{abstract}
Extending previous work [T. Pusztai, L. R\'atkai, A. Sz\'all\'as, and L. Gr\'an\'asy, Phys. Rev. E {\bf 87}, 032402 (2013)], we have studied the formation of eutectic dendrites in a model ternary system within the framework of the phase-field theory. We have mapped out the domain in which two-phase dendritic structures grow. With increasing pulling velocity, the following sequence of growth morphologies is observed: flat front lamellae $\rightarrow$ eutectic colonies $\rightarrow$ eutectic dendrites $\rightarrow$ dendrites with target pattern $\rightarrow$ partitionless dendrites $\rightarrow$ partitionless flat front. We confirm that the two-phase and one-phase dendrites have similar forms, and display a similar scaling of the dendrite tip radius with the interface free energy. It is also found that the possible eutectic patterns include the target pattern, and single- and multiarm spirals, of which the thermal fluctuations choose. The most probable number of spiral arms increases with increasing tip radius and with decreasing kinetic anisotropy. Our numerical simulations confirm that in agreement  with the assumptions of a recent analysis of two-phase dendrites [S. Akamatsu, S. Bottin-Rousseau, G. Faivre, and E. A. Brener, Phys. Rev. Lett. {\bf 112}, 105502 (2014)], the Jackson-Hunt scaling of the eutectic wavelength with pulling velocity is obeyed in the parameter domain explored, and that the natural eutectic wavelength is proportional to the tip radius of the two-phase dendrites. Finally, we find that it is very difficult/virtually impossible to form spiraling two-phase dendrites without anisotropy, an observation that seems to contradict the expectations of Akamatsu {\it et al.}. Yet, it cannot be excluded, that in isotropic systems two-phase dendrites are rare events difficult to observe in simulations.    
\end{abstract}

\maketitle

\section{Introduction}
\begin{figure}[b]
  \includegraphics[width=0.47\textwidth]{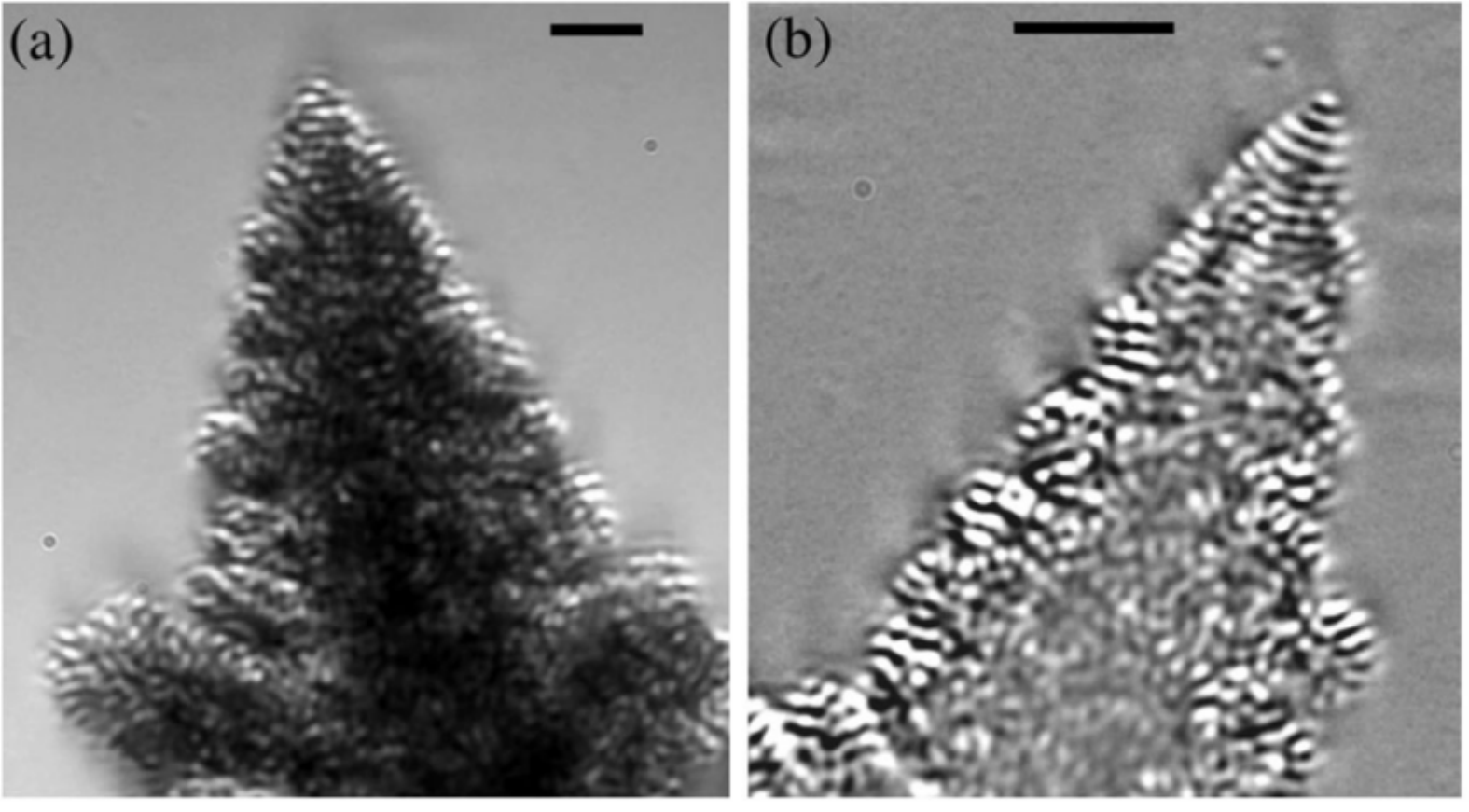}
  \caption{\label{fig:akamatsu} Spiraling eutectic dendrites during bulk-sample directional solidification of a SCN-DC-NA ternary eutectic alloy. Growth rates: (a) $0.28 \: \mu \text{ms} ^{-1}$; (b) $0.95 \: \mu \text{ms} ^{-1}$. The bar corresponds to $20 \: \mu$m. (Reprinted with permission from Ref. \cite{akamatsu_2010} $\copyright$ 2010 American Physical Society.)} 
  
\end{figure}

A broad variety of systems show multi-arm spiraling, including spiral galaxies \cite{galax}, banded spherulites in polymeric systems \cite{band}, biological excitable systems \cite{siegert_1995, siegert_1996}, oscillating chemical reactions \cite{suganthi_2009,thomas_2013a}, spiraling growth edges on flat crystalline surfaces \cite{klemenz_1998}, ridges in sputtered high-temperature superconducting thin films \cite{hawley_1991,gerber_1991}, certain semiconductor materials grown by
molecular beam epitaxy \cite{springholz_1996}, {\it binary} eutectics \cite{fullman_1954, liu_1992}, and helical Liesegang systems \cite{thomas_2013a,thomas_2013b}, and the recently discovered {\it ternary} eutectic dendrites (Fig.~\ref{fig:akamatsu}) \cite{akamatsu_2010}. Although the formation mechanism of spiral structures excite the fantasy of scientists for a long time, a general explanation is not available partly due to the diversity of the underlying physical phenomena. While the details differ in various realizations of spiral growth, diffusion and phase separation often play a role in the respective models. For example, the aggregation of starving cells is controlled by propagating spiral waves of a chemo-attractant, often yielding multiarmed spiral patterns in the case of slime mold \cite{FN}. In {\it binary} eutectics, spiraling has been associated with a specific anisotropy of the free energy of the solid-solid interface \cite{fullman_1954}, the presence of screw dislocations \cite{taran_1976}, or recently to fingering driven by osmotic flow \cite{tegze_2012}. In contrast, the newly discovered spiraling {\it ternary} eutectic dendrites emerge from the interplay of two-phase solidification with the Mullins-Sekerka-type diffusional 
instability caused by a third component, whose solubility differs in the solid and liquid phases \cite{akamatsu_2010}. The complex microstructures observed in some of the ternary alloys \cite{souza_2005} are also suspected to emerge due to the presence of two-phase dendrites \cite{akamatsu_2010}. It is worth noting that besides their scientific interest and pleasing view the latter class of spiraling / helical structures have been identified as possible means of creating chiral metamaterials for optical applications via eutectic self-organization \cite{pawlak1,pawlak2,pawlak3,pendry}.

Models of spiral growth range from wave theory \cite{DW,hakim_1999}, via the FitzHugh-Nagumo (FN) theory for excitable media \cite{FN} and reaction-diffusion models \cite{karma_1999,thomas_2013b,hagberg_1994} to the Ginzburg-Landau/phase-field type models \cite{GL,PF}. Studies of biological excitable media, relying on the FN  model, have clarified essential features of spiral growth \cite{FN}: The mechanism by which the multi-arm spirals do form is the attraction of single spirals rotating in the same direction, whereas the number of spiral arms is associated with the ratio of the single spiral period to the refractoriness of the medium. Apparently, in such systems the multi-arm spiral structures are unstable unless the excitability is sufficiently low. Some other models provide single spiral structures exclusively \cite{GL,PF}. Remarkably, even pure confined systems were found to display chiral symmetry breaking, and thus spiral growth in phase-field simulations \cite{DGK2013}. A recent work based on Cahn-Hilliard type reaction-diffusion model of helical Liesegang systems indicate that single and multiple helixes may occur and the fluctuations choose from the possible configurations \cite{thomas_2013b}. A similar scenario has been outlined in a recent work of us in which ternary eutectic dendrites may realize a variety of eutectic patterns including a target pattern and multiple spiraling motifs of which thermal fluctuations of the system choose \cite{pusztai_2013}. In our work,  we have been looking for a suitable physical model and shown that a simple ternary extension of the standard binary phase-field theory is able to describe the formation of eutectic dendrites, offering thus the first model that is able to capture details of such exotic growth patterns. We note, furthermore, that this is another remarkable success of the phase-field theory: Without any specific adjustments to make spiraling eutectic dendrites, a minimal and fairly standard version of the phase-field theory of ternary solidification proved capable of recovering these rather complex solidification morphologies. We have shown that (a) in many respects eutectic dendrites behave like the single-phase dendrites (e.g., shape, tip radius vs. interface free energy, etc.); (b) there is a multiplicity of eutectic patterns, which lead to steady-state dendritic growth under nominally the same conditions; these are the target pattern, and single- to multi-arm spiraling structures; (c) it is the fluctuations that choose from these competing growth modes; (d) the number of spiral arms tends to increase with increasing dendrite tip radius, however, with some stochastic scattering. In a more recent analytic study, Akamatsu and coworkers \cite{akamatsu_2014} have addressed dendrite formation in ternary eutectic systems. They argued that assuming (i) $\lambda \sim R$ and (ii) $\lambda^2v =$ const. (the latter is the Jackson-Hunt relationship of eutectic solidification \cite{jackson_1966,trivedi_1987}, where $\lambda$, $R$, and $v$ are the eutectic wavelength, the tip radius of the dendrite, and the tip velocity, respectively), the dendrite selection mechanism differs from the one seen in binary systems suggesting that in the case of ternary spiraling dendrites anisotropy is not neccessary for steady-state dendritic growth. Remarkably, assumption (ii) seems to be in contradiction with the analytic predictions by Liu {\it et al.} \cite{liu_2009} for higher concentrations $(5\%)$ of the third component, obtained assuming a rotational paraboloid shape for the two-phase dendrite. However, the approach of Liu {\it et al.} appears to contain questionable assumptions that the lamella spacing is much smaller than the tip radius (expected true rather in the case of eutectic colonies \cite{akamatsu_2000}), and the neglection of the effect of curvature of the dendrite surface on eutectic solidification. It is yet desirable to clarify these contradictory issues.  

Herein, we focus our attention to the ternary eutectic dendrites and extend our previous phase-field study \cite{pusztai_2013} in the following directions: (a) a detailed investigation of the stochastic eutectic pattern selection mechanism; (b) testing of scaling laws (i) and (ii); and (c) interaction of neighboring eutectic dendrites. The structure of our paper is as follows: In Section II, we give a brief description of our model and justify the model parameters used during the simulations. In Section III, we review details of the performed numerical simulations, whereas in Section IV we present our results for the formation of two-phase dendrites, the scaling laws (i) and (ii), the pattern selection mechanism, the role played by the anisotropies, and the dendrite-dendrite interaction. Finally, we summarize our results in Section V and offer a few concluding remarks.

\section{Model description}
We adopt here a simple ternary phase-field model used in our previous study \cite{pusztai_2013}. It can be obtained as a straightforward generalization \cite{plapp_2002,pusztai_2013} of the phase-field model of binary solidification\cite{karma_1994,elder_1994,WB_1995,drolet_2000}, whose variants and extensions have been used successfully for describing homogeneous \cite{granasy_2002,toth_2009} and heterogeneous \cite{granasy_2007,warren_2009,toth_2011} nucleation, and crystal growth including dendritic \cite{WB_1995} and eutectic freezing \cite{karma_1994,elder_1994,drolet_2000}, and even polycrystalline solidification treating nucleation and growth on equal footing \cite{granasy_2002,granasy_2003,granasy_2004,pusztai_2005,pusztai_2008}. Such models turned out to be quantitative on the nanoscale for nucleation \cite{granasy_2002,toth_2009,toth_2011}, whereas qualitative results were obtained for broader interfaces \cite{WB_1995,granasy_2003,granasy_2004,pusztai_2005,pusztai_2008}. A similar model has been used to describe the formation of eutectic colonies in ternary systems \cite{plapp_2002}. 

Herein, the ternary model is solved in a dimensionless form; accordingly, the size scale is determined by the interface thickness chosen for redimensionalizing the results. Considering that the physical interface thickness is about 1-2 nm, the present results refer to nano-scale eutectic structures, expected to form at extreme undercoolings, which can be realized by e.g., via crystallizing amorphous alloys (see e.g., \cite{koster_1975,liu_2008,zhang_2009}). In contrast, the experiments refer to small undercoolings, where other models may be considered more appropriate (e.g.,\cite{karma_2001,folch_2003,folch_2005,plapp_2007}).   

\begin{figure}[t]
  \includegraphics[width=0.47\textwidth]{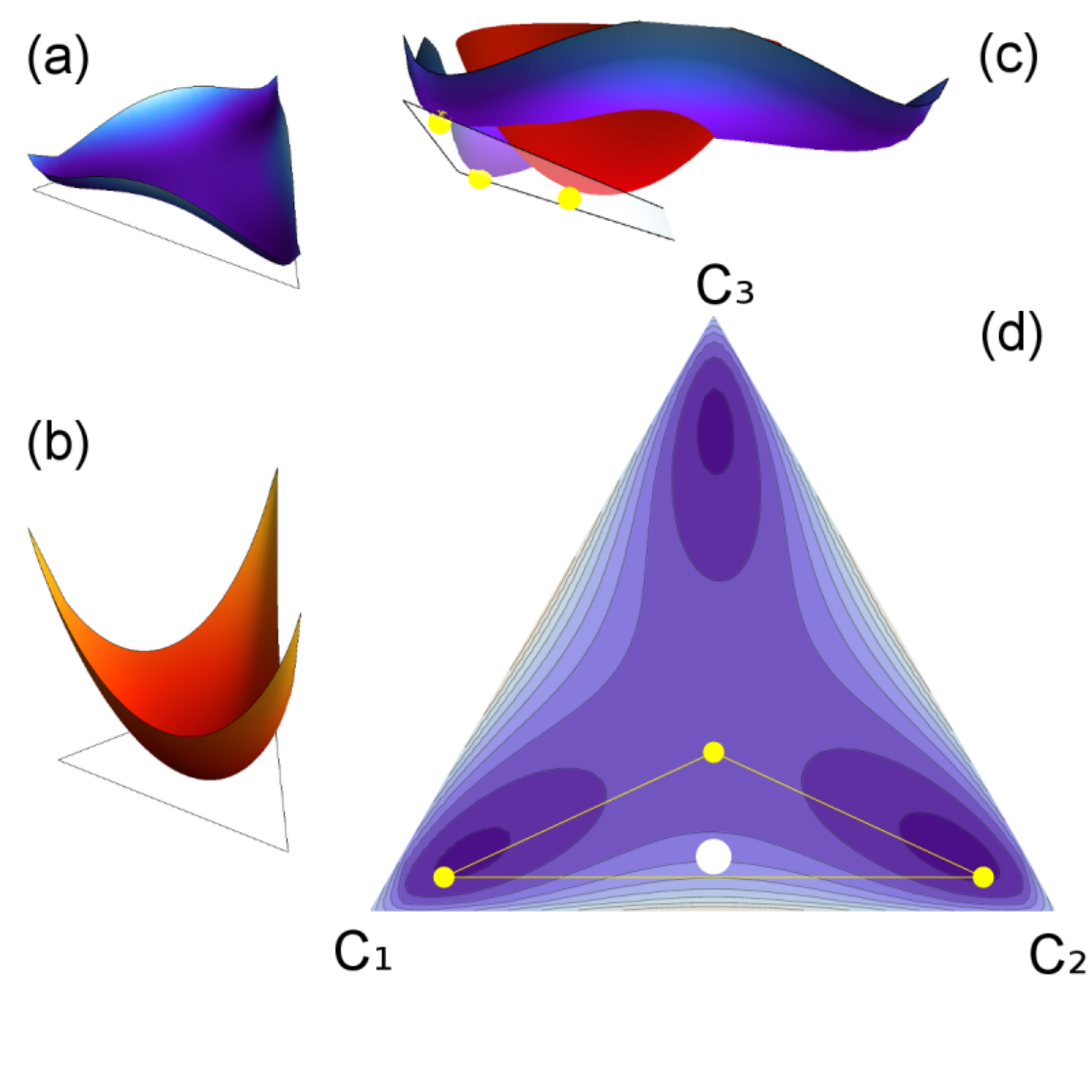}
  \caption{\label{fig:free_energy} The free energy functions, $f_{s,l}(\mathbf{c})$, at a temperature displayed above the Gibbs simplex: (a) bulk solid; (b) bulk liquid; (c) liquid and solid free energy functions with the tangential plane which specifies the equilibrium solid and liquid phases; (d) Yellow circles stand for the three equilibrium phases, the $c_1$- and $c_2$-rich solid phases and the liquid phase. The white circle indicates the composition of the initial liquid typically used in the simulations.}
\end{figure}

The free energy functional has been obtained as a simple generalization of the standard binary phase-field model (see e.g. Refs. \cite{WB_1995, granasy_2002}):
\begin{eqnarray}
F[\phi,\mathbf{c}] = & \int \left[ \frac{\epsilon_{\phi}^2}{2} (\nabla\phi)^2 + w g(\phi) +  (1-p(\phi))f_l(\textbf{c}) + \right. \nonumber \\
& \left.   + p(\phi) \left( f_s(\textbf{c}) + \frac{\epsilon_c^2}{2} \sum_{i=1}^{3} (\nabla {c_i})^2 \right) \right] \, dV.
\end{eqnarray}
Here $\phi(\mathbf{r}, t) \in [0, 1]$ is the phase field, which is 0 for bulk liquid and 1 for solid, $\mathbf{c}=(c_1, c_2, c_3)$ are the concentration fields, the quartic function $g(\phi)=\phi^2(1-\phi)^2/4$ ensures the double-well form of $F$, while function $p(\phi)=\phi^3(10-15\phi+6\phi^2)$ switches on and off the free energy densities of the solid and liquid phases. We choose ideal solution model for the bulk liquid phase [Fig.~\ref{fig:free_energy}(b)]:
\begin{eqnarray}
f_{l}(\mathbf{c}) = \sum_{i=1}^3 c_i \left[f^{l}_{i} + \log c_i \right],
\end{eqnarray}
where $f_i^l=0$, and the regular solution model for the solid [Fig.~\ref{fig:free_energy}(a)]:
\begin{eqnarray}
f_{s}(\mathbf{c}) = \sum_{i=1}^3 c_i \left[f^{s}_{i} + \log c_i \right] + \frac{1}{2}\sum_{i,j, i \ne j} \Omega_{ij} c_i c_j.
\end{eqnarray}
The equations of motion (EOMs) derived variationally have the following forms for the phase field
\begin{eqnarray}
\dot{\phi} &= \epsilon (\mathbf{n}) M_{\phi} \left[ \epsilon_{\phi}^2 \nabla^2 \phi - w g^\prime(\phi) + \right. \nonumber \\
&\left. + p^{\prime}(\phi)(f_l(\mathbf{c}) - f_s(\mathbf{c})) - p^{\prime}(\phi) \frac{\epsilon_c^2}{2} \sum_{i=1}^{3} (\nabla {c_i})^2, \right]
\end{eqnarray}
and for the concentration fields 
\begin{eqnarray}
\dot{c_i} = \sum_{j=1}^{3}\nabla \cdot \left[(1-p(\phi))M^c_{i,j}\left(\nabla\frac{\delta{F}}{\delta{c_j}}\right)\right],
\label{eq:cdot}
\end{eqnarray}
where $\delta F/\delta c_i$ is the functional derivative of the free energy with respect to concentration field $c_i$. Here the $\sum_i c_i = 1$ constraint is automatically satisfied by our choice of 1 and $-0.5$ for the diagonal and off-diagonal elements of the 3 $\times$ 3 mobility matrix, $\mathbf{M^{c}}$. Note that diffusion is switched off in the bulk solid. Owing to this and the ideal solution thermodynamics assumed in the bulk liquid, phase separation is possible in only the solid-liquid interface layer. 

In a few test cases that explore the effect of concentration fluctuations, we included a conservative (flux) noise for the concentration fields, via adding $\nabla \mathbf{\Xi}_{i}$ to the RHS of the respective EOMs [Eqs. (\ref{eq:cdot})], where the vector $\mathbf{\Xi}_{i}(\mathbf{r},t)$ is a random current of component $i$. Since local mass conservation ($\sum_i c_i = 1$) has to be obeyed even in the presence of fluctuations, the sum of the divergencies of these random currents must be zero, i.e., $\sum_i \nabla \mathbf{\Xi}_i = 0$ has to be satisfied. Local mass conservation is realized by the construction $\mathbf{\Xi}_{i} = \sum_{i \ne j} \mathbf{\xi}_{ij}$, where $\mathbf{\xi}_{ij}$ represents that part of the total current $\mathbf{\Xi}_{i}$, where the flow of component $i$ is compensated by the back-flow of component $j$ as dictated by the linked term $\mathbf{\xi}_{ij}(\mathbf{r},t)=-\mathbf{\xi}_{ji}(\mathbf{r},t)$ in the respective expression for $\mathbf{\Xi}_{j}$. When solving the discretized equations, the random currents $\mathbf{\xi}_{ij}=-\mathbf{\xi}_{ji}$  were realized by exchanging random amounts of components $i$ and $j$ between each possible pairs of neighboring cells and for all the $i$--$j$ combinations. Random numbers of Gaussian distribution and zero mean were used to determine the amount of the components exchanged.

We have considered in this work a kinetic anisotropy of cubic symmetry, represented by the following orientation dependent coefficient in Eq. (4):
\begin{equation}
 \epsilon (\textbf{n}) = (1-3\epsilon_4) \left[1 + \frac{4\epsilon_4}{1-3\epsilon_4} (n_x^4 +n_y^4 +n_z^4)\right].
\end{equation}
Here $\textbf{n} = \nabla \phi/|\nabla \phi|$, whereas parameter $\epsilon_4$ determines the magnitude of anisotropy. As this anisotropy acts at the solid-liquid interface independently of composition, the same kinetic anisotropy applies for the $\alpha$--liquid and $\beta$--liquid interfaces, as well as to the two-phase-solid--liquid interface. Owing to neglecting the anisotropy of the solid-liquid interface free energy, the present computations are regarded as qualitative.

\section{Numerical simulations}
The equations of motion were solved numerically in a dimensionless form using the finite difference method and explicit time stepping. The simulations were performed in a massively parallel manner partly on a computer cluster of 928 CPU cores and on several graphics processing unit (GPU) cards housed at the Wigner Research Centre for Physics, Budapest, Hungary.  In a few cases, where extreme large simulations were required, we used the supercomputer Hitachi Super Technical Server SR16000 Model M1 of the Center for Computational Materials Science, at the Institute of Materials Research, Tohoku University, Sendai, Japan. 

\begin{figure}[b]
\includegraphics[width=0.47\textwidth]{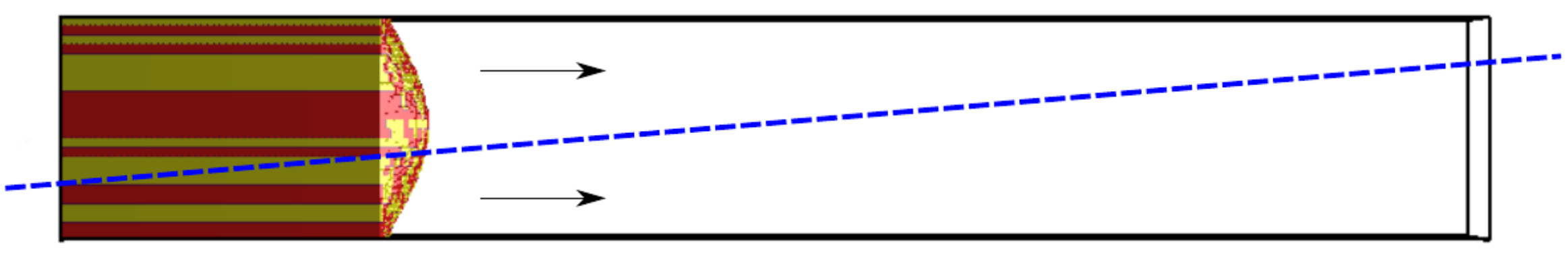}
  \caption{\label{fig:simbox} Lateral view of the simulation box. Grid size: $N_x \times N_y \times N_z$ = $96 \times 96 \times 612$, unless stated otherwise. The dashed line indicates the temperature distribution. The yellow and red domains stand for the solid phases, while the liquid is made transparent.}
\end{figure}

The simulations have been carried out under conditions that correspond to directional solidification.
A temperature gradient was imposed via making the free energy of the solid dependent on position: $f_{s,i,\tilde{z}} = f_{s,i}^{(0)}-\tilde{z}\left(\partial{f_{s,i}} / \partial{\tilde{z}}\right)$. Here $\tilde{z}$ is the spatial coordinate along the direction of the pulling velocity $\tilde{v}_p$. In order to model the pulling of the sample, the contents of the arrays $\phi$ and $c_i$ have been shifted by one voxel back in the $\tilde{z}$ direction in each $[(d\tilde{x}/d\tilde{t})/\tilde{v}_p]$\textsuperscript{th} time step, while prescribing the  boundary conditions $\phi=0$ and $\mathbf{c} = \mathbf{c_0}$ on the high $T$ and no-flux boundary conditions on the low $T$ side of the simulation box.

To make simulations that are sufficiently large in the direction of pulling ($\tilde{z}$) more efficient, only Eqs. ({\ref{eq:cdot}}) have been solved far ahead of the solidification front, where the phase field is small ($< 10^{-8}$) to be regarded as bulk liquid. This approach allowed us to handle essentially the whole diffusion zone ahead of the solidification front, which then ensured that the dendrite morphology became independent of the length of the simulation box ($N_z$).

As in our previous study \cite{pusztai_2013}, the majority of our simulations has been started by including a slab of solid of length $(2/9)\tilde{L}_z$ into the simulation box ($\tilde{L}_z = N_z d\tilde{x}$ is the length of the simulation box in direction $\tilde{z}$) with a small hump at the center, whereas the initial composition of the solid has been set to $\langle c_1 \rangle = \langle c_2 \rangle = 0.455$ and $c_3 = 0.09$ [see Fig.~\ref{fig:free_energy}(d)], realized by a random transversal ($\tilde{x}-\tilde{y}$ plane) distribution of the two solid phases of typical size scale close to the natural wavelength of eutectic growth. The random transversal distribution represents here the effect of all the thermal fluctuations from the nucleation of the eutectic grain to the starting of the simulation. We have opted for this initial condition, because simulations that follow the formation of the two-phase dendrites from fluctuation-induced emergence of nuclei are prohibitively time consuming. Evidently, this might be a crude approximation. Therefore, we tested in a few cases how close one can get to the results obtained from this manner if we start the solidification from a small but supercritical (larger than the critical size for nucleation) one-phase solid sphere, while adding flux-noise to the EOMs of the concentration fields as described in Ref. \cite{echebarria}. As will be seen, the probabilities of formation of the individual eutectic patterns obtained via these two routes are fairly close, indicating physical consistency.    

{\it Reference conditions:} These were used unless stated otherwise. Time and spatial steps: $d\tilde{t} = 0.0025$ and $d\tilde{x} = 1.0$.
Grid: $N_x \times N_y \times N_z$ = $96 \times 96 \times 612$.
Size of the simulation box: $\tilde{L}_i = N_i \cdot d\tilde{x}$, where $i = x, y,$ or $z$.
Composition: $c_1 = c_2=0.455$, and $c_3 = 0.09$.
Parameters of free energy densities: $f_{l,i}=0$; $f_{s,i}^{(0)}=-0.9640$ and $\partial{f_{s,i}}/\partial{\tilde{z}}=3.677 \times 10^{-4}$. At $\tilde{t} = 9000$ (steady-state), this yields $f_{s,i} = -0.9140$ at the dendrite tip, corresponding to a rather substantial relative undercooling of $\Delta \tilde{T}_r = (\tilde{T}_L-\tilde{T})/(\tilde{T}_L-\tilde{T}_S) \approx 0.78$, where $\tilde{T}_L$ and $\tilde{T}_S$ are the dimensionless liquidus and solidus temperatures corresponding to the conditions at the dendrite tip.
Pulling velocity: $\tilde{v}_p=0.2$, whereas $\Omega_{1,2} = 3.05$,  $\Omega_{2,3} = \Omega_{3,1} = 3.0$. $\tilde{M}_{\phi} = 1.0.$
Solid-liquid interface free energy (isotropic): $\tilde{\gamma}_{SL,0} = 0.0147$.
Kinetic anisotropy: $\epsilon_4 = 0.3$. Coefficients of the square-gradient terms: $\epsilon_{\phi}^2 = 0.75, \epsilon_{c}^2 = 0.4,$ and the free energy scale, $w = 0.0469$. Unless stated otherwise no noise has been added to the EOMs.

\begin{figure}[t]
  \includegraphics[width=0.45\textwidth]{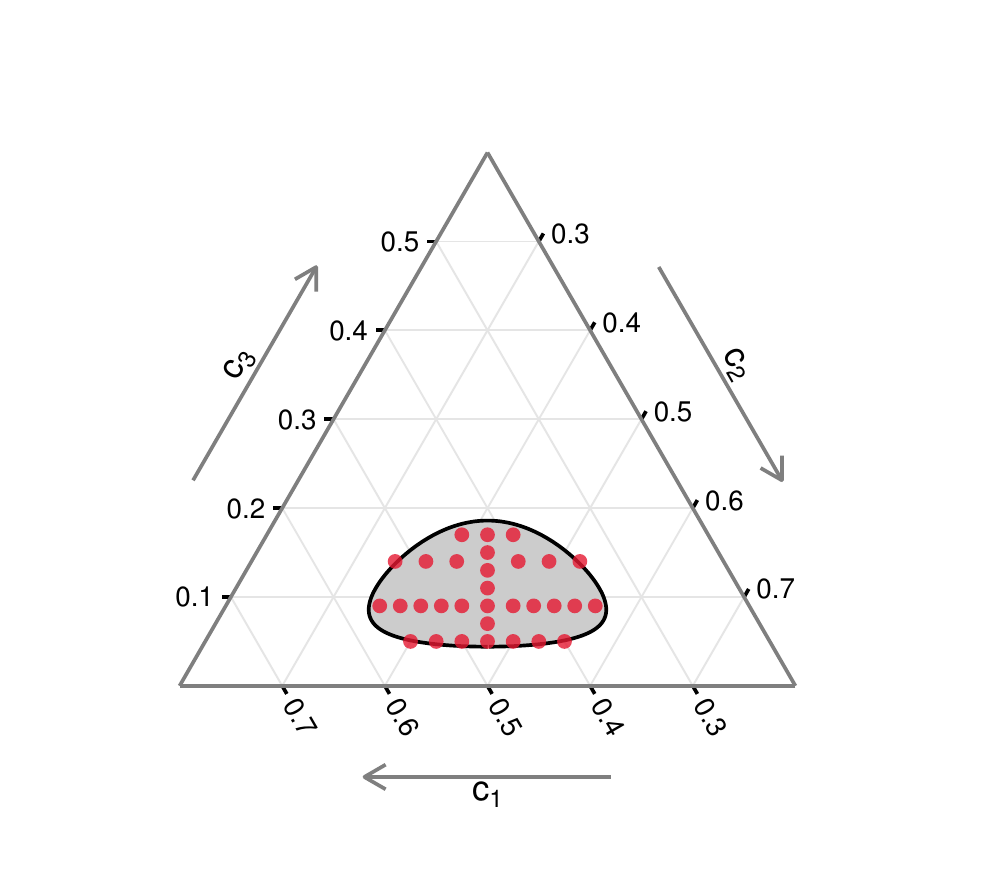}
  \caption{\label{fig:mapping} Compositions at which two-phase dendrites of target or spiral patterns have been observed at $\tilde{v}_p = 0.2$ (red dots). }
\end{figure}

\begin{figure*}[t]
  \includegraphics[width=0.9\textwidth]{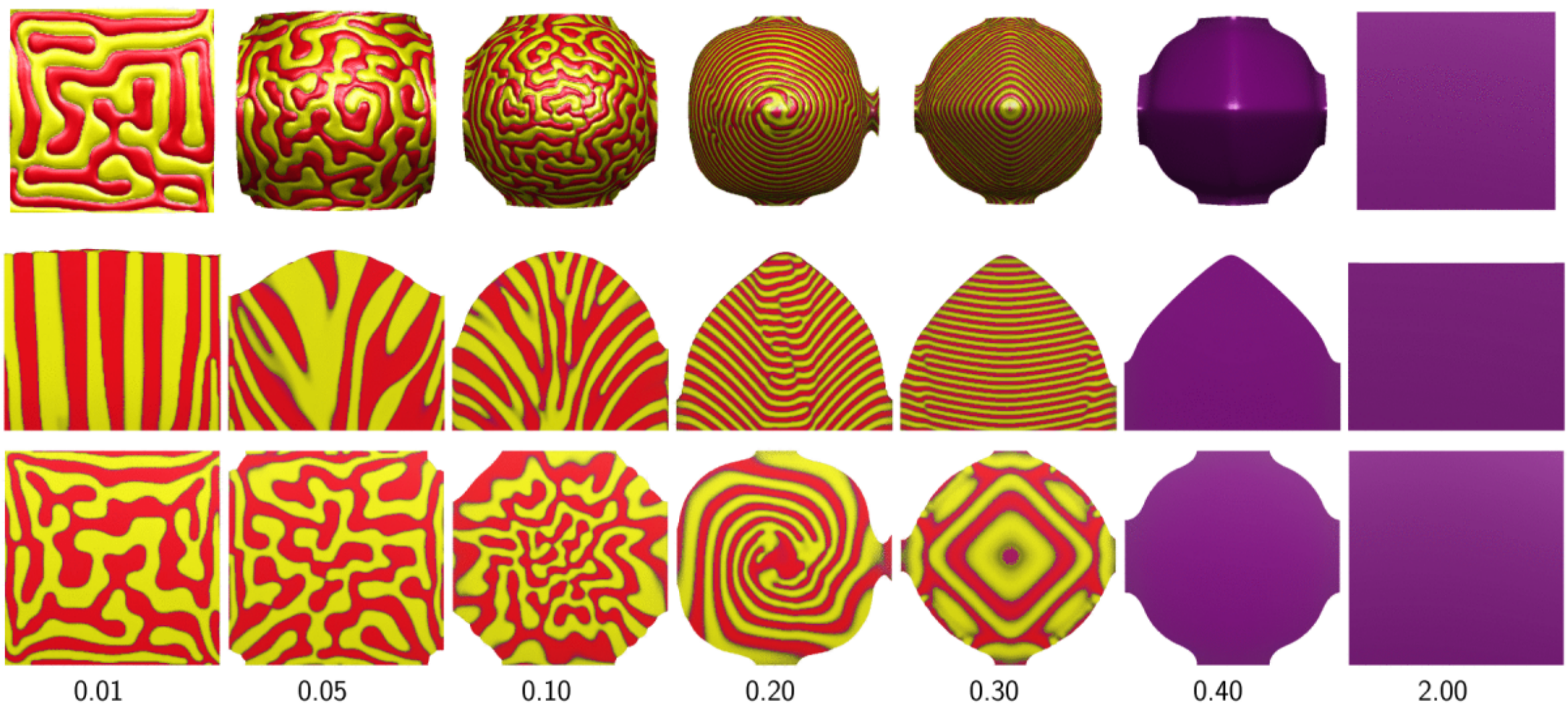}
  \caption{\label{fig:vpull} Solidification morphology and pattern formation as a function of dimensionless pulling velocity $\tilde{v}_p$. The $c_1$ and $c_2$-rich solid solutions are colored red and yellow, respectively, whereas the liquid is transparent, and purple stands for $c_1 \approx c_2$. The front view (top row), the longitudinal section (central row) and transverse cross sections (bottom row) are displayed. With increasing velocity, the sequence of growth morphologies is: flat front lamellae $\rightarrow$ eutectic colonies $\rightarrow$ eutectic dendrites $\rightarrow$ dendrites with target pattern $\rightarrow$ partitionless dendrites $\rightarrow$ partitionless flat front. At extremely low and high velocities planar fronts develop. Above $\tilde{v}_p \approx 0.35$ solidification takes place without partitioning. At extreme high velocities ($\tilde{v}_p \gtrsim 0.8$) full solute trapping occurs. Remarkable is the gradual transition from the (usual) lamellae parallel to the temperature gradient seen at low velocities to lamellae  perpendicular to the temperature gradient seen at $\tilde{v}_p \approx 0.3$.}
\end{figure*}

\begin{figure}[b]
  \includegraphics[width=0.45\textwidth]{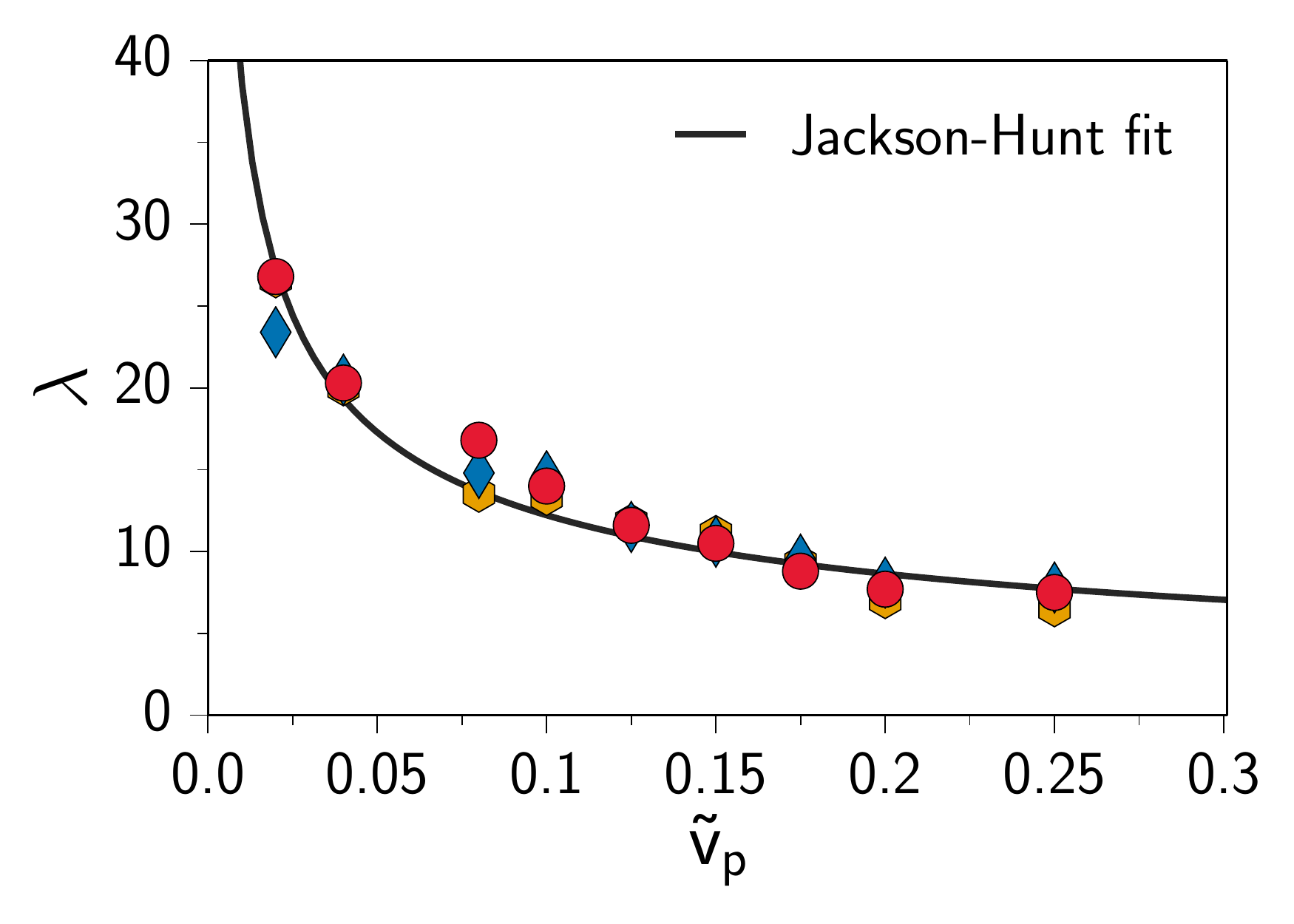}
  \caption{\label{fig:JH_scaling} The wavelength of eutectic solidification as a function of pulling velocity. (Symbols of different colors denote series of simulations with different initial patterns.) Note the reasonable agreement with the Jackson-Hunt scaling, $\lambda \propto \tilde{v}_p^{-1/2}$ (solid line).}
\end{figure}

\begin{figure}[t]
  \includegraphics[width=0.45\textwidth]{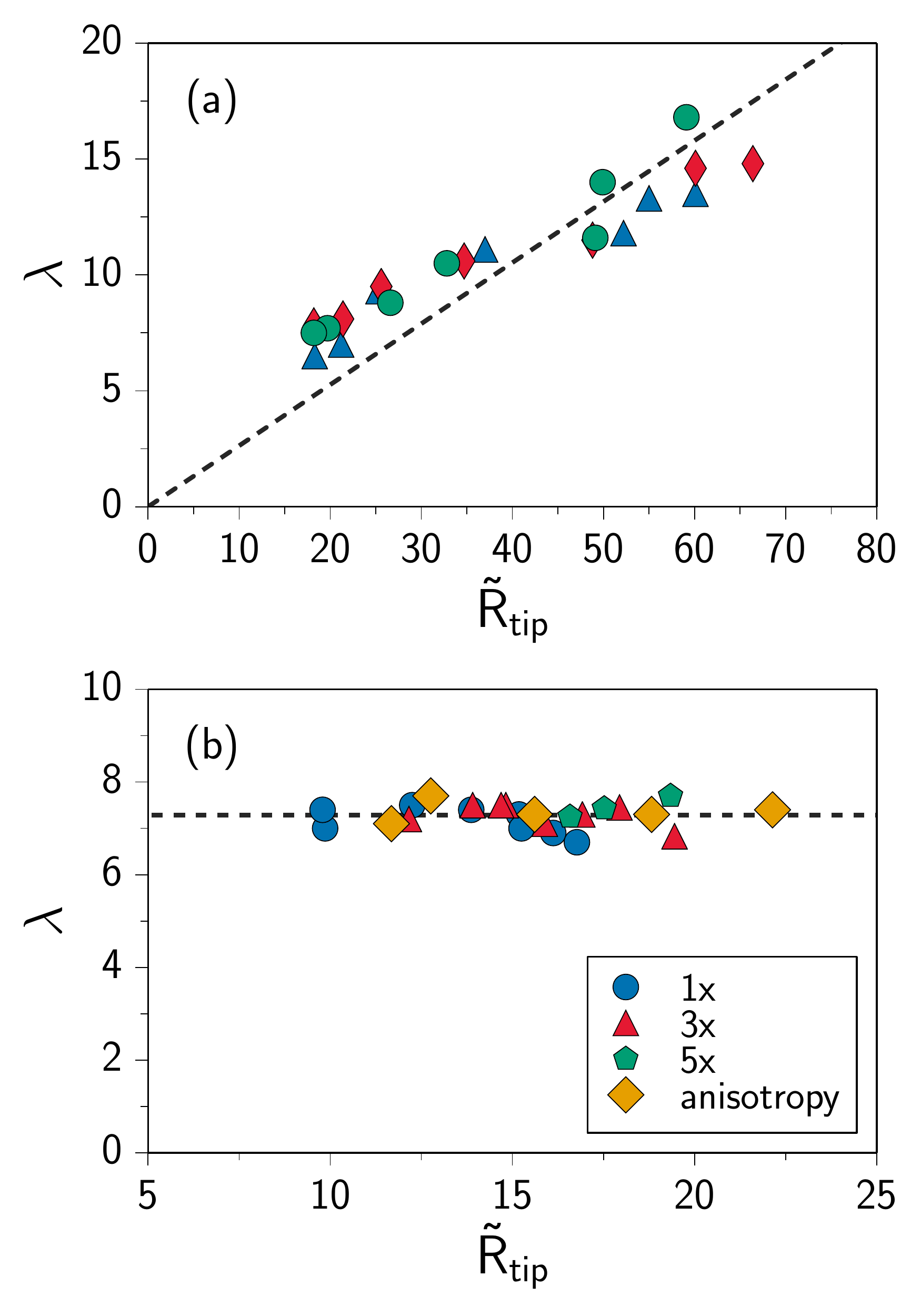}
  \caption{\label{fig:scal2} Eutectic wavelength vs. tip radius. 
(a) Velocity is varied, while other parameters are fixed. Different symbols stand for runs with different initial patterns (different initializations of the random number generator). (b) Varying the interface free energy $\gamma_{SL,0}$ or the kinetic anisotropy $\epsilon_4$ while keeping the velocity constant. (Here circles, triangles and pentagons indicate the number of spiral arms observed in simulations where $\gamma_{SL,0}$ was varied, while diamonds stand for results from simulations where the kinetic anisotropy $\epsilon_4$ was varied. In the experiments $\gamma_{SL,0}$ and $\epsilon_4$ are fixed, under such conditions the two characteristic lengths are roughly proportional to each other, $\lambda \propto R_{tip}$, as assumed in Ref. \cite{akamatsu_2014}. However, the tip radius can be influenced by varying either the magnitude of the interfacial free energy or the kinetic anisotropy. The dashed lines are to guide the eye.} \end{figure}

\section{Results and Discussion}
In a previous work on spiraling eutectic dendrites \cite{pusztai_2013}, we explored the parameter space defined by composition, temperature gradient, pulling velocity, interfacial free energy, and kinetic anisotropy, and optimized the conditions for growing two-phase steady-state dendritic structures. Herein, we extend these studies to significantly broader parameter ranges. We examine, furthermore, the relationship between the tip radius and the eutectic wavelength, investigate the validity of Jackson-Hunt scaling ($\lambda^2v =$ const. \cite{jackson_1966,trivedi_1987}), present a detailed characterization of the morphology of the forming eutectic structures, including the 3D geometry of the dendrite tip, and the internal distribution of the constituent phases, and investigate whether anisotropy is indeed not needed for the appearance of two-phase dendrites.      

\subsection{Domain of ordered two-phase dendrites}
To identify the region, in which spiraling eutectic dendrites do form in the phase diagram, we have investigated the solidification morphology in the neighborhood of the operating point [$\bf{c}$ = (0.455, 0.455, 0.09) and $\tilde{v}_p =  0.2$], reported in Ref. \cite{pusztai_2013}. The respective domain is shown in  Fig.~\ref{fig:mapping}, which indicates that spiraling eutectic dendrites may be observed in an oval region that extends to $0.35 < c_1 < 0.56$ along the line $c_3 = 0.09$, whereas it covers the range $0.05 < c_3 < 0.17$ along the line of symmetric compositions ($c_1 = c_2$). 

\begin{figure}[b]
  \includegraphics[width=0.47\textwidth]{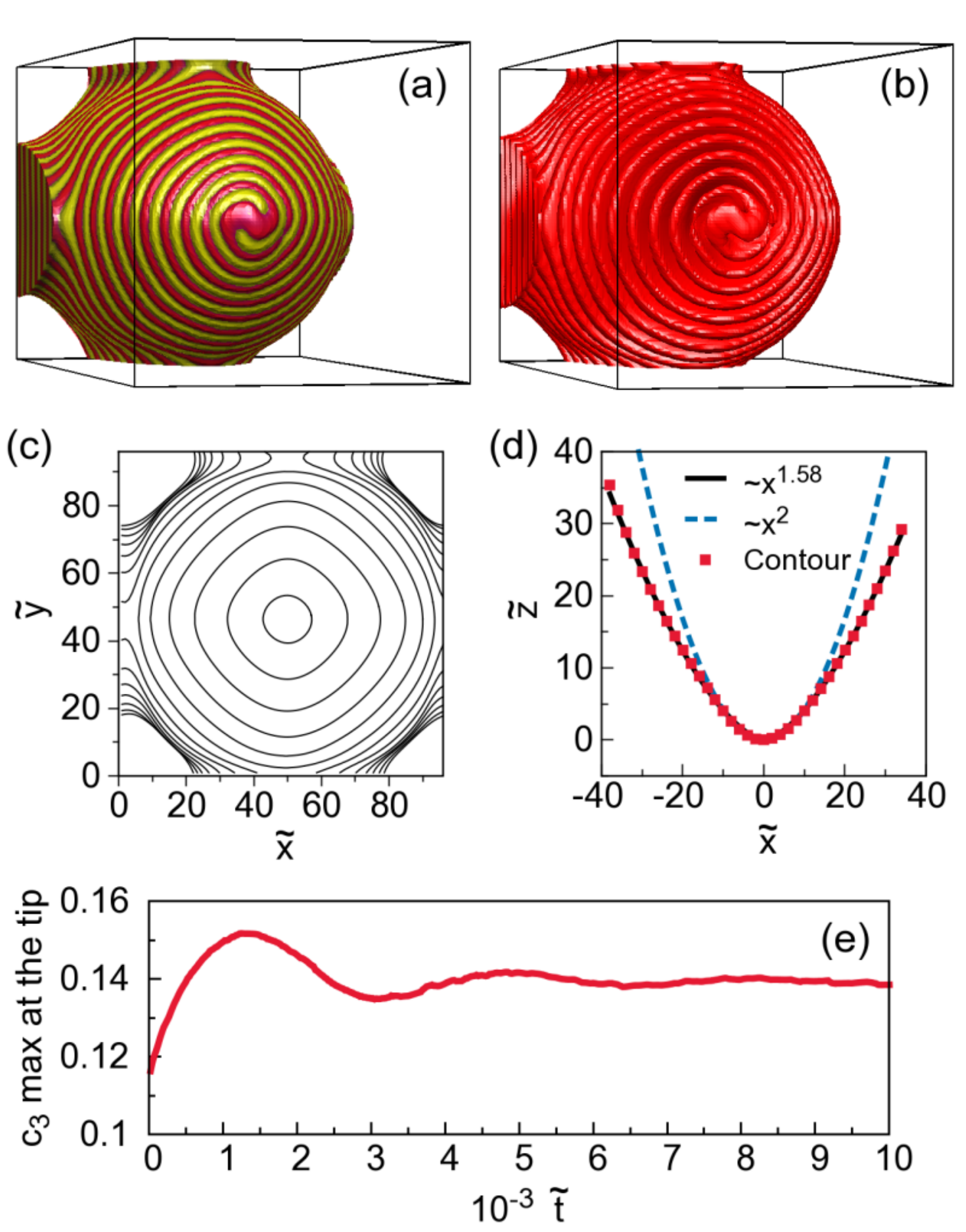}
  \caption{\label{fig:double_spiral} Two-phase two-armed spiraling dendrite grown under the reference conditions defined in Section III: (a) spiraling motif on the surface; (b) the helical structure formed by one of the solid phases, (c) contour lines showing the transverse sections at $10 d\tilde{x}$ distances; (d) longitudinal section (dots), the best fit parabola (dashed), and the curve $\tilde{z} = \tilde{z}_{max}-|\tilde{x}|^{\nu}$ fitted to it (solid line); (e) maximum of $c_3$ at the tip vs time.}
\end{figure}

\subsection{Jackson-Hunt scaling}   
To see whether the $\lambda^2 v =$ const. relationship applies indeed, we have investigated the solidification morphology as a function of pulling velocity at the composition $\bf{c}$ = (0.455, 0.455, 0.09). For low pulling velocities ($\tilde{v}_p < 0.01$) one finds a planar front with a disordered lamellar pattern (see Fig.~\ref{fig:vpull}). Increasing $\tilde{v}_p$ resulted in a dendritic morphology yet with disordered lamellar pattern. Around $\tilde{v}_p \sim 0.2$ ordered eutectic patterns form such as the target and spiraling patterns. At higher $\tilde{v}_p \approx 0.3$, solidification takes place via forming lamellae perpendicular to the temperature gradient. Above this velocity partitionless dendrite forms indicating that solute trapping does not occur for $c_3$ at the same level as for the other two components. At even higher pulling velocities ($\tilde{v}_p \approx  2.0$) partitionless growth with a flat interface has been observed. The results shown in Fig.~\ref{fig:vpull} suggest that the transition from flat lamellar eutectic structure to the partitionless growth with flat interface with increasing velocity happens via the following sequence: flat lamellar front $\rightarrow$ eutectic colonies $\rightarrow$ eutectic dendrites $\rightarrow$ dendrites with target pattern $\rightarrow$ partitionless dendrites $\rightarrow$ partitionless flat front. Remarkably, during the transition from a flat surface with lamellae perpendicular to it (seen at small velocities) to a flat surface at full solute trapping (at high velocities), happens via a stage preceding solute trapping, in which a nearly 1D eutectic pattern evolves: lamellae that are perpendicular to the temperature gradient. Similar transition (lamellae parallel with the temperature gradient to lamellae perpendicular to it) has been observed when increasing the velocity in binary eutectic simulations \cite{elder_1994,drolet_2000,toth_unpub}. 

Apparently, the natural length scale of eutectic solidification follows the Jackson-Hunt scaling ({\it cf.} symbols and solid line in Fig.~\ref{fig:JH_scaling}), confirming thus the assumption made in Ref. \cite{akamatsu_2014}. The other assumption of Ref. \cite{akamatsu_2014} that the tip radius and the eutectic wavelength are proportional $\lambda \propto R_{tip}$  is satisfied likewise provided that the other parameters are kept constant  [see Fig.~\ref{fig:scal2}(a)]. In our case, the proportionality coefficient $\lambda = (0.26 \pm 4.5\%) R_{tip}$  is much smaller than in the experiments $\lambda \approx 1.33 R_{tip}$ \cite{akamatsu_2014}. One can however, tune the the ratio via changing the interface free energy or the kinetic anisotropy:  If the velocity is kept constant and either the interface free energy is increased or the kinetic anisotropy decreased, the tip radius increases, whereas the eutectic wavelength remains roughly constant [Fig.~\ref{fig:scal2}(b)]\cite{lambconst}. Accordingly, one could move towards the experimental conditions via decreasing the interface free energy or increasing the kinetic anisotropy. Presumably, the presence of anisotropic interface free energy could also be used to tune $\lambda/R_{tip}$, however, its investigation is left for the future.

\subsection{Dendrite morphology}
A typical two-phase dendrite is shown in  Figs.~\ref{fig:double_spiral}(a) and \ref{fig:double_spiral}(b). It has a rounded square-like transverse section in the $\tilde{x}-\tilde{y}$ plane [Fig.~\ref{fig:double_spiral}(c)]; whereas in the fin directions (e.g., $\tilde{x}-\tilde{z}$), the longitudinal profile can be fitted reasonably well by $\tilde{z} = \tilde{z}_{max}-|\tilde{x}|^{\nu}$, where $\tilde{z}_{max}$ is the tip position, $\tilde{x}$ is the distance from the axis of the dendrite, and $\nu$ is $\sim 1.58 \pm 0.05$ [Fig.~\ref{fig:double_spiral}(d)]. This exponent is somewhat lower than $\nu = 5/3$ predicted theoretically for single-phase dendrites \cite{brener}, or the $\nu = 1.67$ obtained experimentally for xenon dendrites \cite{bisang_1995_xenon, bisang_1996,singer_2004}. (In our work, the perimeter of the dendrite has been defined as the contour line $\phi = 0.5$.) The quasi steady growth form (not a true steady-state as the eutectic pattern rotates in the coordinate system moving with the dendrite tip) has been established after a transient characterized by decaying oscillations of the tip-radius, tip-temperature, and the maximum of $c_3$ at the tip, as shown in Fig.~\ref{fig:double_spiral}(e). These findings are in a reasonable agreement with a detailed characterization of tip shapes of single-phase dendrites by phase-field simulations \cite{karma_2000}, which indicates that there is a crossover between a corrected paraboloid valid close to the tip and the xenon-like behavior on the tail; therefore, the results depend (here too) somewhat on the range of distances one choses to fit the shape. 

\begin{figure}[t]
  \includegraphics[width=0.45\textwidth]{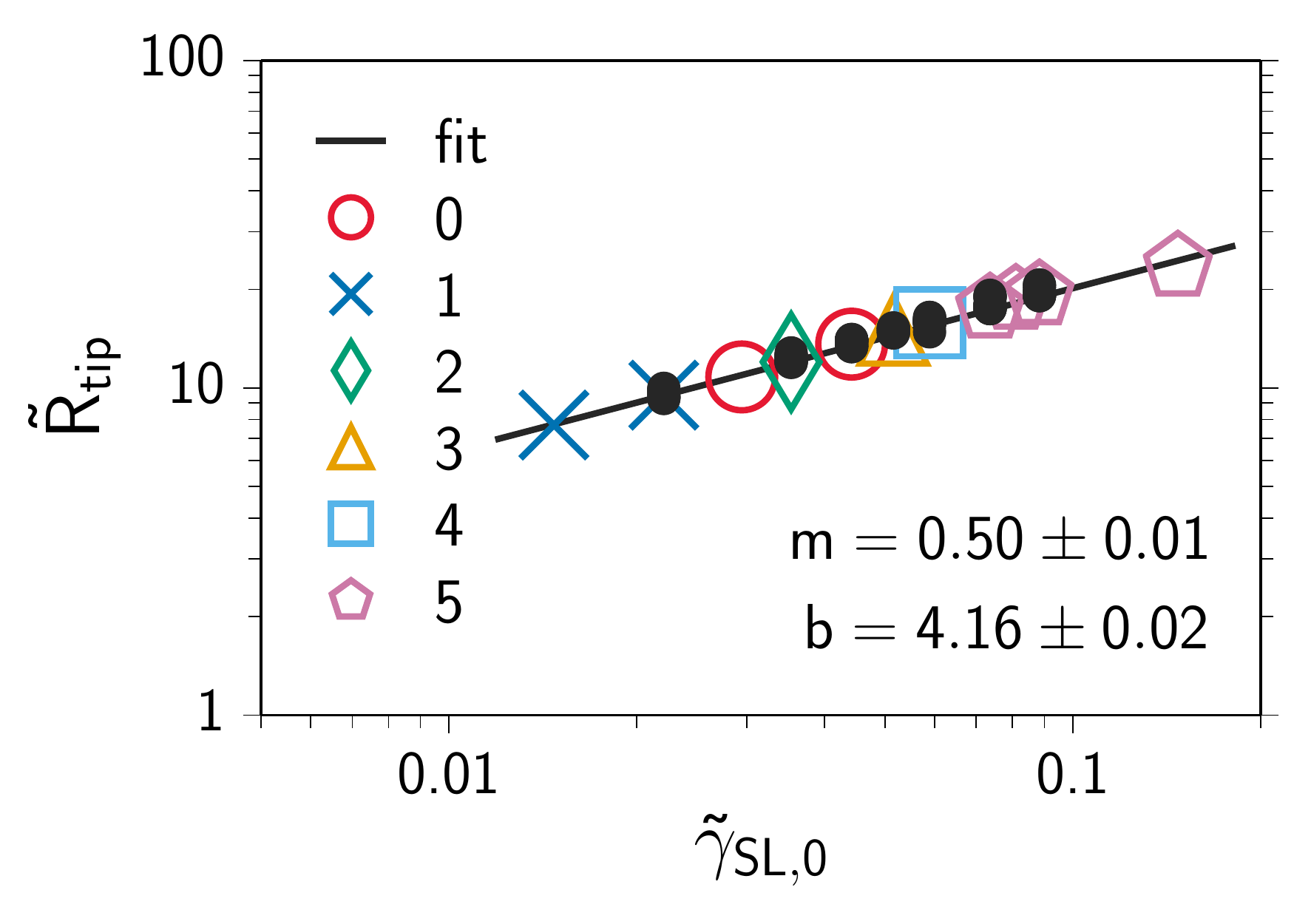}
  \caption{\label{fig:gsl_vs_r} Tip radius vs. solid-liquid interface energy. Results obtained with the same initialization but at different $\tilde{\gamma}_{SL,0}$) values are depicted by different empty symbols indicating the actual eutectic pattern. In contrast, the black symbols stand for 20 additional simulations, which were performed under the same condition as the underlying empty ones, except that the initial random spatial distribution of the solid phases was different (achieved by usig different initializations for the random number generator). These simulations realize various eutectic patterns, yet they fall on the same master curve, indicating that the shape of the dendrite is essentially independent of the eutectic pattern.} 
\end{figure}

To see how far the two-phase dendrites follow the behavior of  single-phase dendrites, we have varied the magnitude of the solid-liquid interface free energy ($\tilde{\gamma}_{SL}$) via changing the free energy of the single component solid-liquid interface ($\tilde{\gamma}_{SL,0}$). After the dendrite reached its quasi steady-state, we evaluated and averaged the tip radius in the fin directions ($\tilde{R}_{tip}$). The results are summarized in Fig.~\ref{fig:gsl_vs_r}. It was found that $\tilde{R}_{tip} \propto \tilde{\gamma}_{SL,0}^{0.50 \pm 0.01}$, which is very close to the theoretical relationship $\tilde{R}_{tip} \propto \tilde{\gamma}_{SL}^{1/2}$ derived for single-phase dendrites (see e.g., \cite{kurz_fisher_1989} or \cite{dantzig_rappaz_2009}. It probably indicates that the chemical contribution is negligible to the interfacial free energy $\tilde{\gamma}_{SL}$. We should also call attention to the fact that the shape of the dendrite (and the tip radius) is independent of the two-phase pattern forming on the surface of the dendrite: The target pattern, and single- and multiple spiraling motifs appear to coexist under the same conditions (for further details see Sub-section IV.D below), yet the probability of finding a larger number of  $N_{\text{arm}}$ increases with increasing tip radius. 

\begin{figure}[t]
  \includegraphics[width=0.4\textwidth]{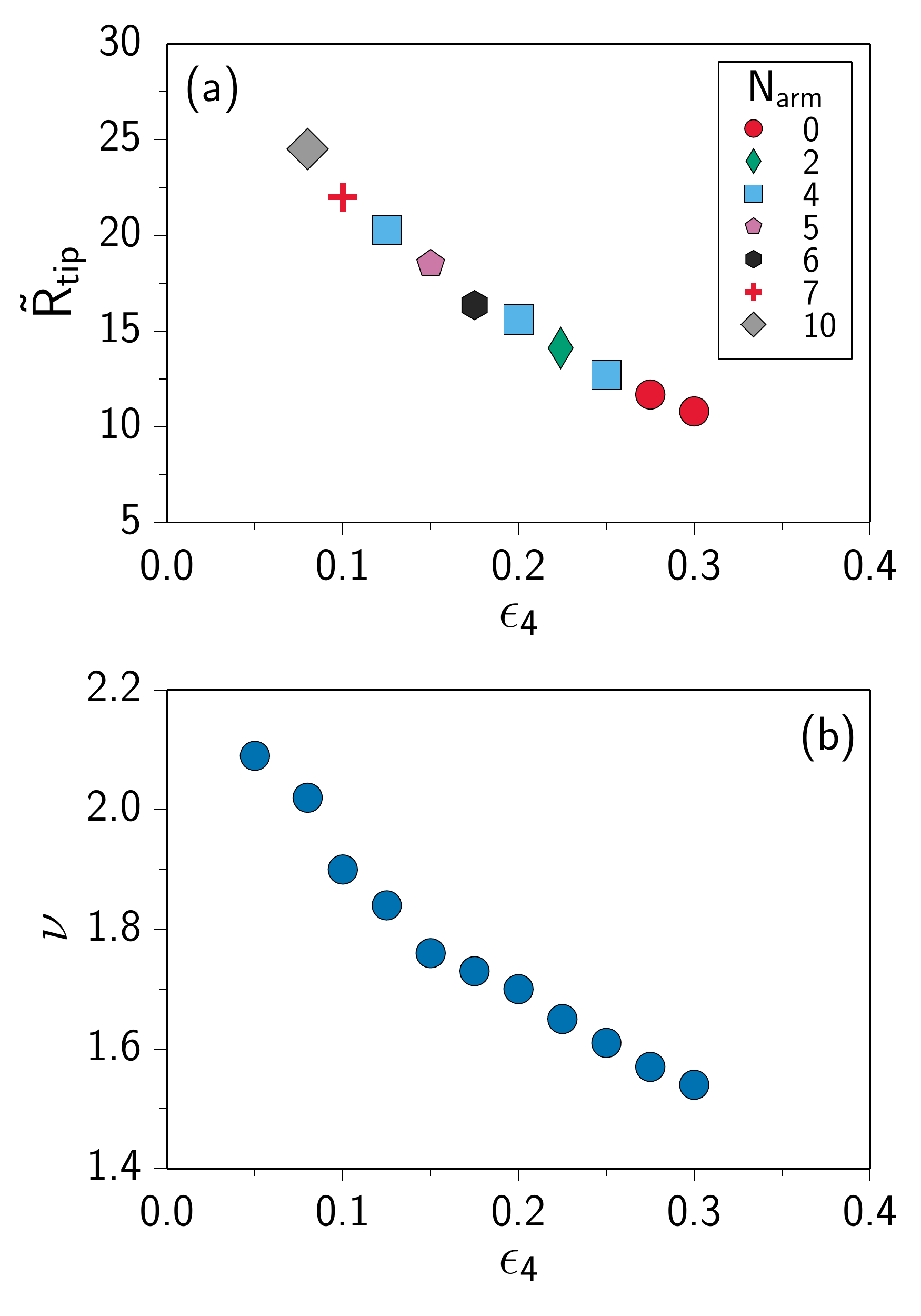}
  \caption{\label{fig:anisotropy} The effect of kinetic anisotropy on the shape of the two-phase dendrite. (a) Tip radius vs. kinetic anisotropy at $\tilde{\gamma}_{SL,0}= 0.0295$; (b) Exponent $\nu$ vs. kinetic anisotropy. ($\nu$ has been obtained by fitting the expression $\tilde{z} = \tilde{z}_{max}-|\tilde{x}|^{\nu}$ to the longitudinal cross section of the dendrite, where $\tilde{z}_{max}$ is the tip position, $\tilde{x}$ is the distance from the axis of the dendrite.)}
\end{figure}

Another way to influence the shape of the dendrites is to change the kinetic anisotropy ($\epsilon_4$). The results are shown in Fig.~\ref{fig:anisotropy}(a): decreasing anisotropy yields an increasing tip radius, a change apparently initiating a larger number of the spiral arms. The variation of the kinetic anisotropy influences the shape of the cross-section in the direction of the fins. The exponent $\nu$ describing the shape of the dendrite tip changes from $\sim 1.5 \pm 0.1$ to $\sim 2.1 \pm 0.1$ with decreasing anisotropy [see Fig.~\ref{fig:anisotropy}(b)]; i.e., it varies roughly between the experimental value for xenon (1.67) and the rotational paraboloid (2.0) expected for isotropic case. This is combined with a change of the shape of transverse section from a square of rounded corners to a circle. Apparently, there appears to be some sort of a correlation between the tip radius and the number of spiral arms: the probability of  having larger number of spiral arms increases with increasing tip radius. We note that the reason for observing $\nu >$ 2 is probably finite-size effect: the influence of the boundary of the simulation box on the dendrite shape exerted far from the tip region.

\subsection{Eutectic patterns}
\begin{figure}[t]
  \includegraphics[width=0.47\textwidth]{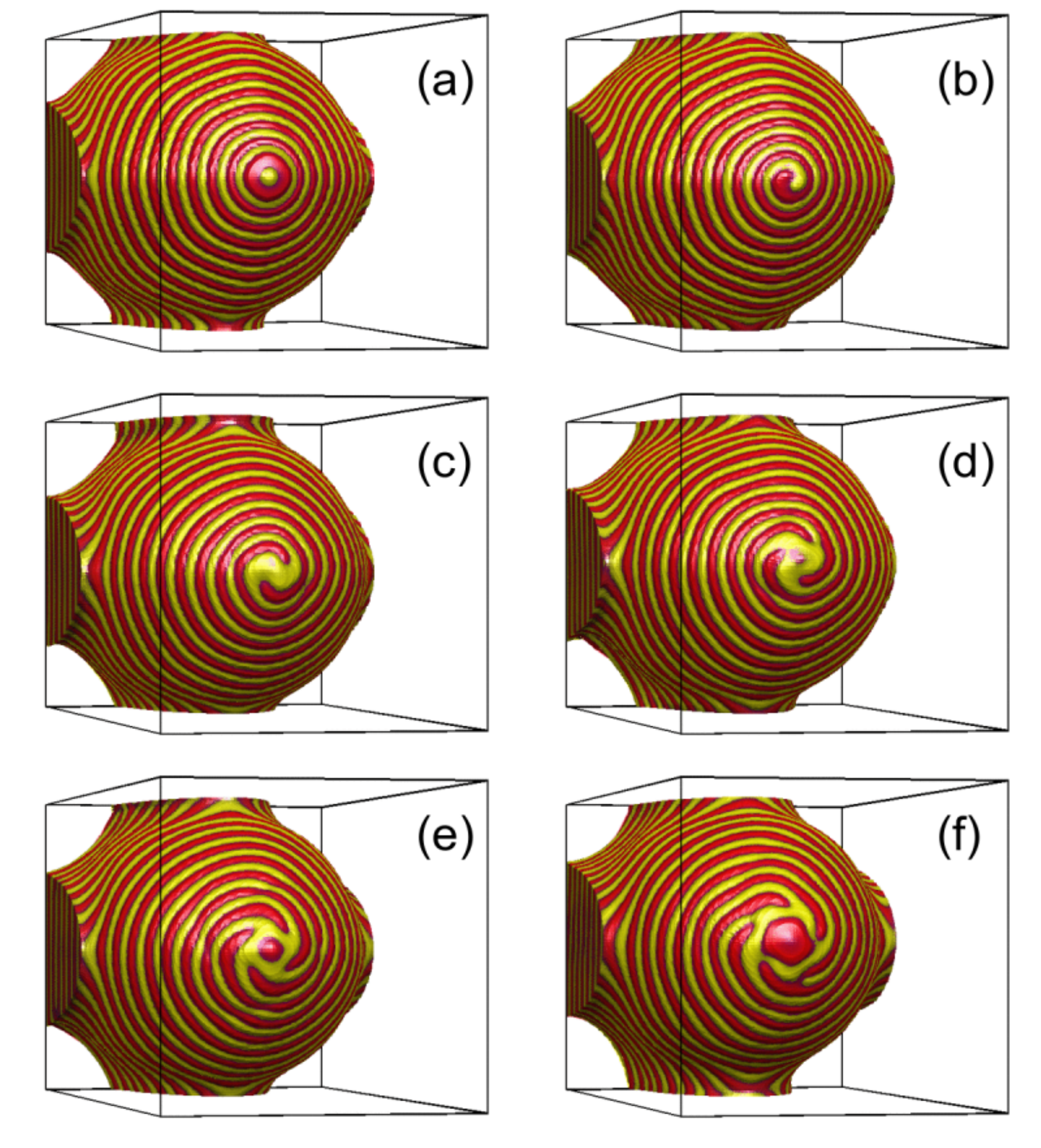}
  \caption{\label{fig:multiarm_spirals} Eutectic patterns observed under nominally the same conditions (reference conditions as specified in Section III): (a) Target and, (b) single- and (c)-(f) multiarm spiral motifs. They differ in only the initialization of the random number generator used in creating the initial random eutectic pattern.}
\end{figure}

Next, we investigate the formation of two-phase patterns on the surface of eutectic dendrites. As reported previously \cite{pusztai_2013} they include target-, single- and multi-arm spiraling motifs (Fig.~\ref{fig:multiarm_spirals}). A closer inspection of the tip region reveals that several modes of  pattern formation are possible. 

The {\it target pattern} appears via alternating nucleation of the two solid phases on top of each other [Fig.~\ref{fig:multiarm_spirals}(a)], confining the occurrence of this mode to larger undercoolings. However, no nucleation is needed for the formation of {\it single- and double spirals}, in which the eutectic pattern originates from alternative occupation of the tip region by the two solid solution phases. In the latter cases this happens so that the $\alpha$ and $\beta$ phases remain spatially continuous: It appears that the alternating occupation of the tip region by the two phases may happen without nucleation. Having a growing $\alpha$ region at the tip, component B piles up ahead of the tip until fast growth initiated by the previously formed $\beta$ phase leads to the formation of a $\beta$ domain at the tip, which is thus connected to the rest of the $\beta$ phase. This is followed by the the same process, but now for the $\alpha$ phase. Since in this way the formation of spiral arms (single or multiple) does not need nucleation, these mechanisms can prevail at small undercoolings. The growth modes with larger number of spirals ($N_{\text{arm}} \geq 3$) become increasingly more complex, still displaying alternating phase appearance at the tip. Owing to the complexity of the process, it is difficult to decide whether nucleation plays a role here or not. The larger is the number of spiraling arms, $N_{\text{arm}}$, the more disordered is the tip region. As a result, larger numbers of point- and line defects travel down on the surface of the dendrite. This is especially so for the simulations performed with flux noise for the concentration fields.

\begin{figure}[t]
  \includegraphics[width=0.5\textwidth]{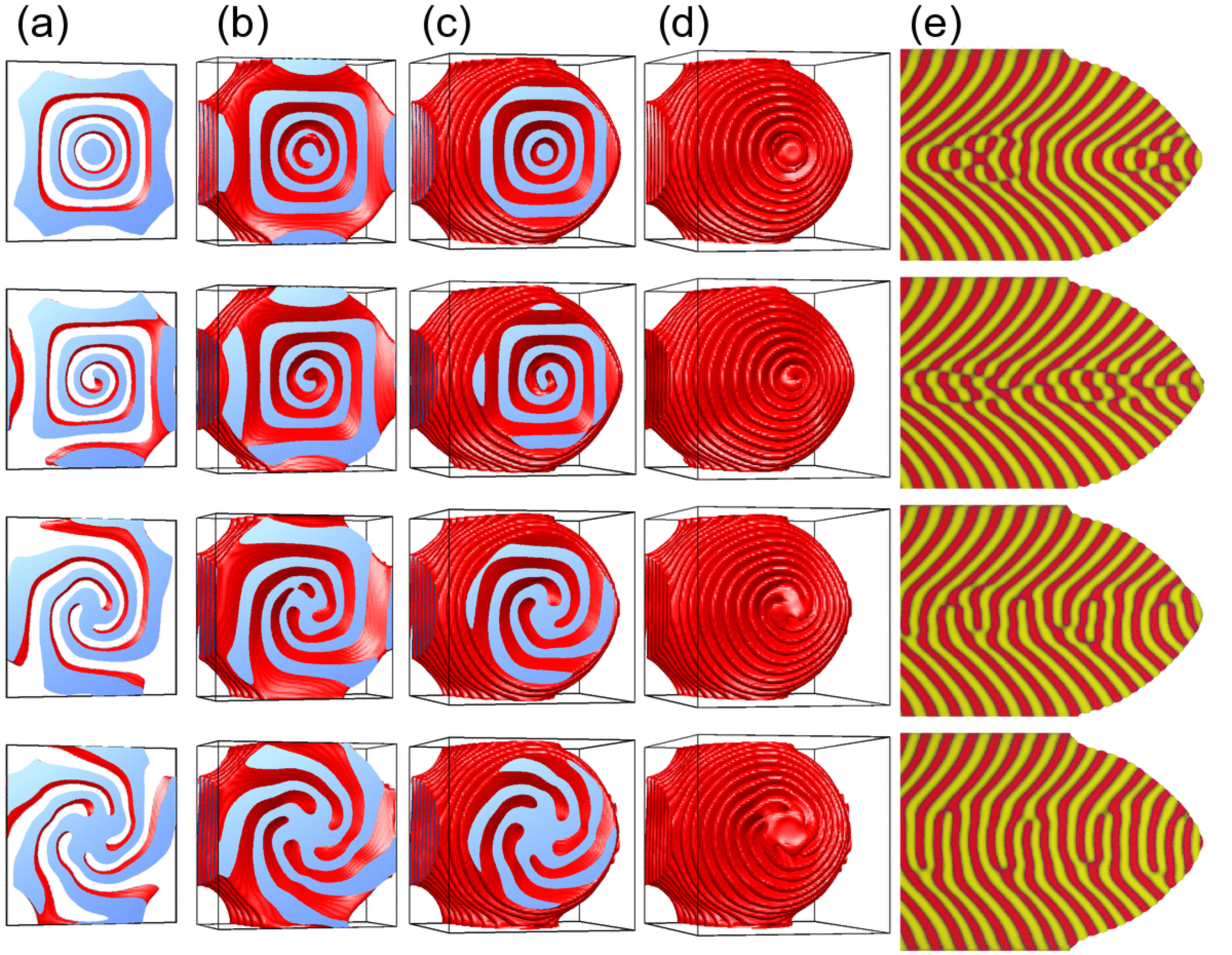}
  \caption{\label{fig:crosssec} Dendrite cross-sections for the target, single-, triple- and five-arm cases shown in Fig.~\ref{fig:multiarm_spirals}. (a)-(c): one phase $x-y$ plane frontal view, the box cut at $z=3, 50 \textnormal{ and } 100$, respectively, perpendicular to $z$ axis. The eutectic structures are connected inside the dendrite. (d) Helical structure formed by one of the phases. The cut surfaces were colored with blue for better contrast. (e) $y-z$ plane cross-section.}
\end{figure}

The internal structure of eutectic dendrites with the target pattern, the single-, triple-, and five-arm spiraling modes are presented in Fig.~\ref{fig:crosssec}. Apart from periodically occurring defect-rich regions, the target pattern is composed of conical domains of the two phases, where the conical domains of the same phase are not connected spatially (i.e., alternating nucleation of the two phases is needed to create them). In contrast, the single- and multiple spiraling  patterns observed on the surface are realized by single- or multiple helical structures forming in the volume. Owing to obvious geometrical constraints, the steepness of the spirals on the surface  increases with the number of spiral arms. Remarkably, the longitudinal sections of the two-phase dendrites are fairly similar for all these modes, although weak systematic differences can be observed [Fig.  \ref{fig:crosssec}(e)]. More characteristic are the front views and the transverse sections: The individual eutectic growth modes (number of spirals) can clearly be distinguished [see Fig.  \ref{fig:crosssec}(a)-(d)]. It has been observed that in the appropriate parameter domain, the spiraling two-phase dendrites are quite robust. It is worth noting that in the experiments performed at low undercoolings the single-spiral mode has been exclusively observed  \cite{akamatsu_2010}. The multiple spiral arms seen in our simulations probably follow from the large relative undercooling we applied.

\begin{figure}[t]
  \includegraphics[width=0.45\textwidth]{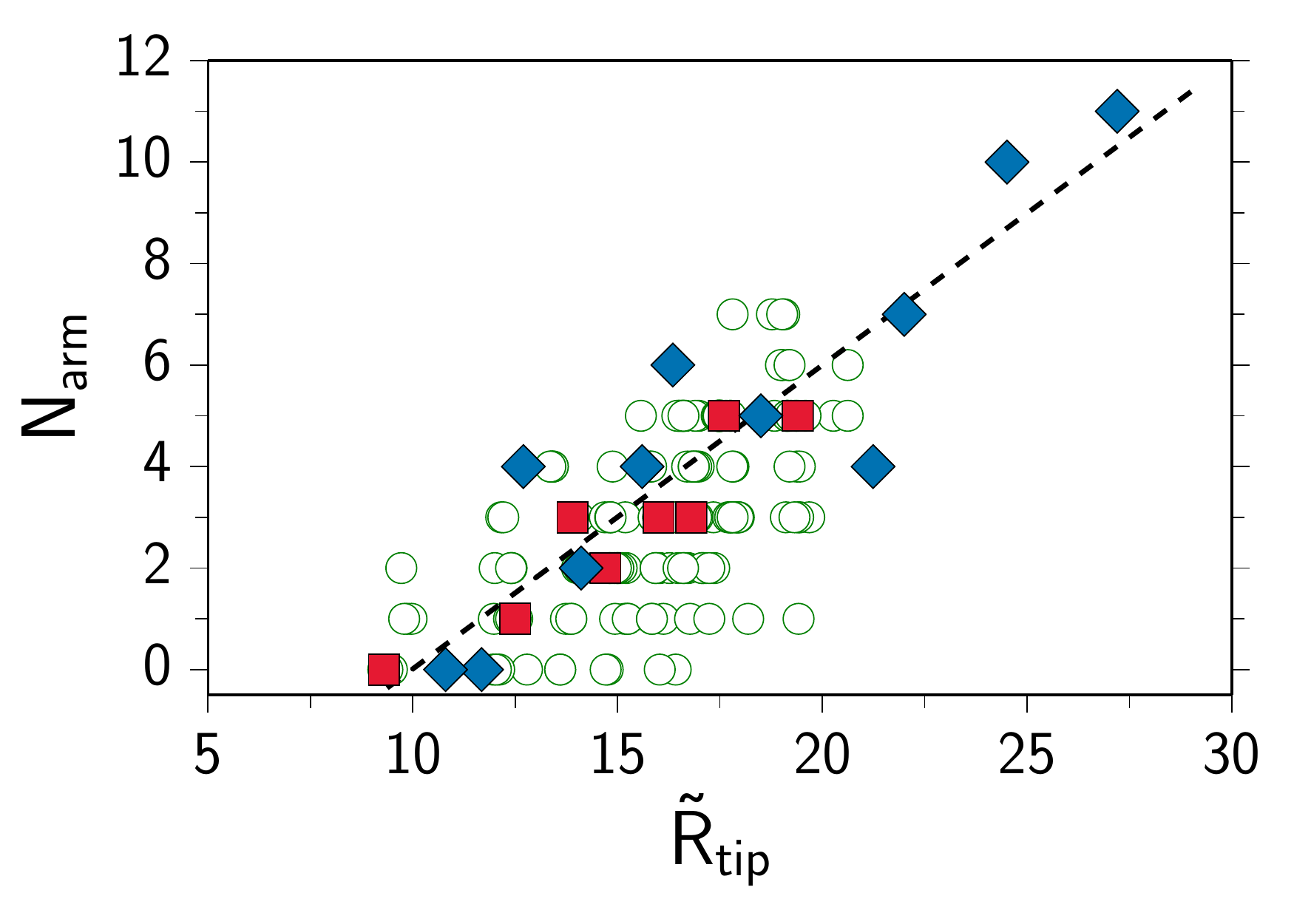}
  \includegraphics[width=0.47\textwidth]{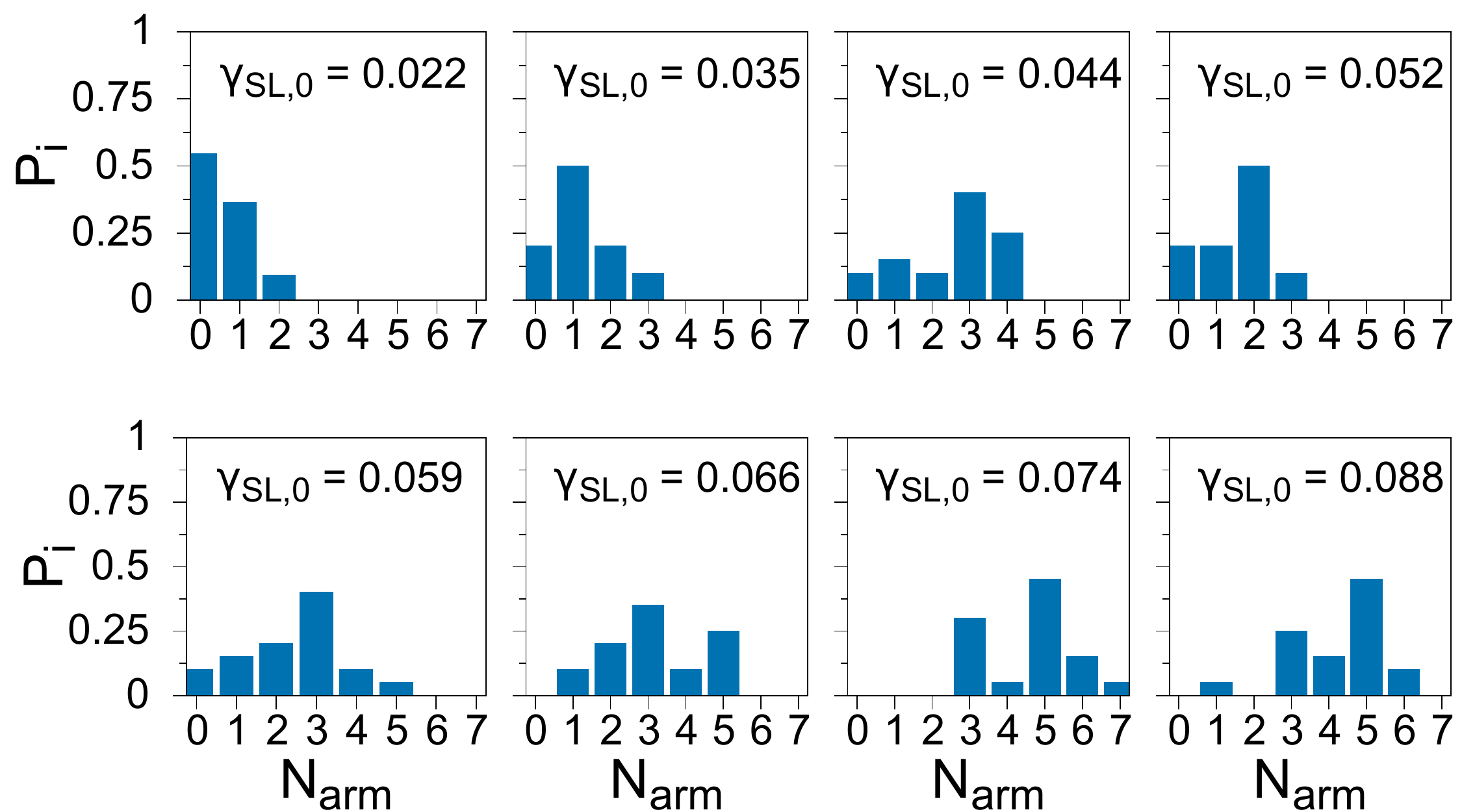}
  \caption{\label{fig:r_vs_narm} Pattern selection for two-phase dendrites: Top: Number of spiral arms vs. tip radius. Empty (green) circles: results from random initial eutectic patterns at $\epsilon_4=0.3$ (180 simulations). Blue diamonds: same initialization, but $\epsilon_4$ varies between $0.3$ and $0.05$. Red squares: the most likely eutectic patterns from 20 random initializations (evaluated from the respective 20 empty circles). Bottom: Probability distributions (histograms) for steady-state eutectic patterns obtained from simulations started from 20 different random initial eutectic patterns.}
\end{figure}

Although the number of spiraling arms ($N_{\text{arm}}$) tends to increase with increasing solid-liquid interfacial free energy (Fig.~\ref{fig:gsl_vs_r}), which in turn is reflected in increasing tip radius, the steady-state pattern depends also on the initial random distribution of the two solid phases representing the cumulative effect of the preceding compositional fluctuations: Different steady-state patterns were obtained starting from different (random) initial patterns (see Fig.~\ref{fig:r_vs_narm}). For example at $\tilde{\gamma}_{SL,0} = 0.059$, these patterns include the target pattern together with single- to fivefold spirals, which indicates the multiplicity of the possible steady-state solutions for the same physical conditions of which the random initialization chooses.

This stochastic behavior is characterized by a peaked probability distribution (see central and bottom rows in Fig.~\ref{fig:r_vs_narm}), which shows increasing number of spiral arms with increasing solid-liquid interfacial free energy (and therefore tip radius). 

\begin{figure}[t]
  \includegraphics[width=0.45\textwidth]{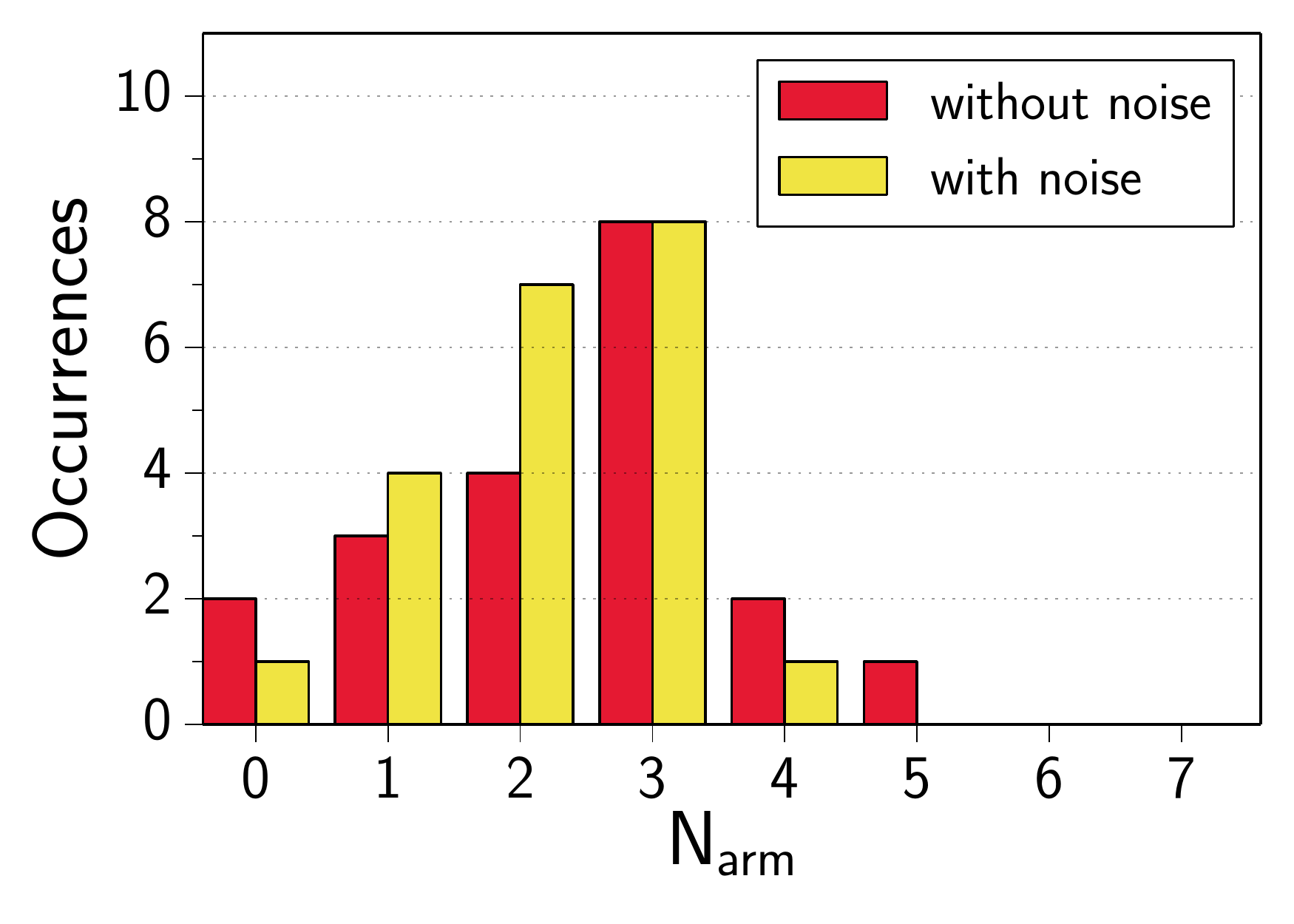}
  \caption{\label{fig:rnd_test} Probability distributions (histograms) for the accessible eutectic patterns differing in the number of spiral arms obtained on the basis of 20 simulations. Red: started from random initial eutectic patterns without noise. Yellow: started from a single-phase seed while applying flux noise in the EOMs for the concentration fields. The individual runs differed in the initialization of the random number generator. Note the similarity of the results obtained with and without flux noise for the concentration fields.}
\end{figure}

In order to check the validity of the assumption that such random initialization can reasonably represent the effect of compositional fluctuations, we have performed 20 simulations starting from a single-phase crystal seed while adding flux noise to the EOMs for the concentration fields as in Ref. \cite{echebarria}. These simulations differed in only the initialization of the random number generator. Such simulations are rather time consuming since growth from the seed to the steady-state dendrite has to be covered, and the adding of flux noise also slows down the simulation process. The corresponding probability distribution is compared to the distribution obtained from computations with random initial two-phase patterns in Fig.~\ref{fig:rnd_test}. Considering the scattering of the results due the relatively small number of simulations limited by the available computational power, a reasonable agreement can be seen between the two types of probability distributions. 

Apparently, in our case the thermal fluctuations decide which steady-state solutions are accessible for the system for a given set of operating parameters. These features closely resemble the behavior reported for Liesegang patterns, where the fluctuations determine, which of the competing modes (rings, single-, double-, triple- and multiple helices, or disordered patterns) is realized \cite{thomas_2013a,thomas_2013b}. These findings raise the possibility that stochastic pattern selection is universal for certain multiarm spiral systems, in which diffusion plays a decisive role. Further work is needed, however, to clarify this issue.

\begin{figure*}[t]
  \includegraphics[width=0.75\textwidth]{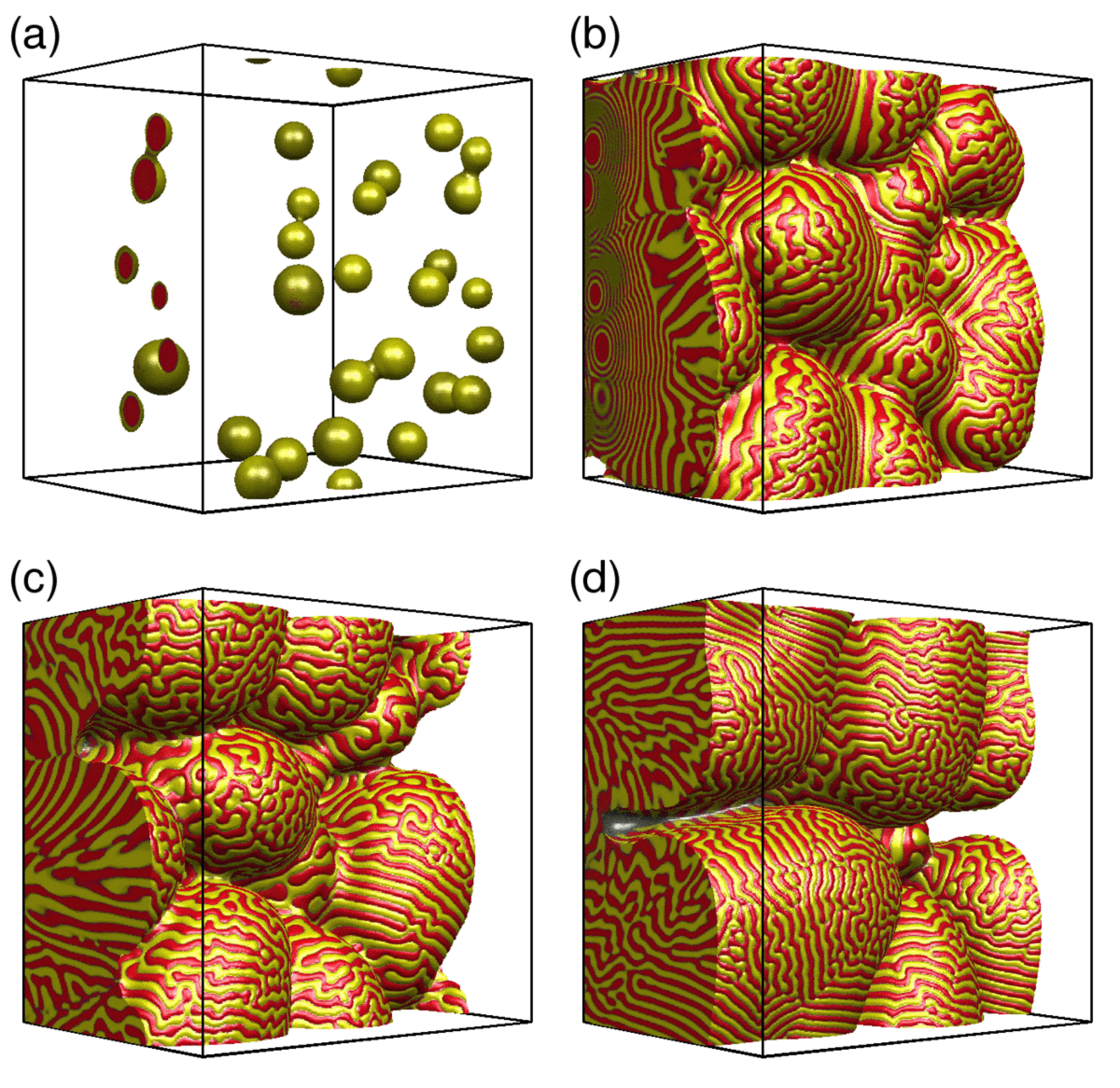}
  \caption{\label{fig:large} Formation of eutectic colonies in the model ternary eutectic system. Snapshots of a large scale simulation ($380 \times 380 \times 511$ grid) performed at $v_p = 0.1$ without anisotropy (taken at $10^4, 6.2\times10^5, 1.6\times10^6,$ and $4.0\times10^6$ time steps). Here solidification was started from 25 single-phase seeds. Note the competing cellular structures evolving after a transient period, and that the cellular structure dynamically changes with time.}
\end{figure*}

\begin{figure*}[t]
  \includegraphics[width=0.75\textwidth]{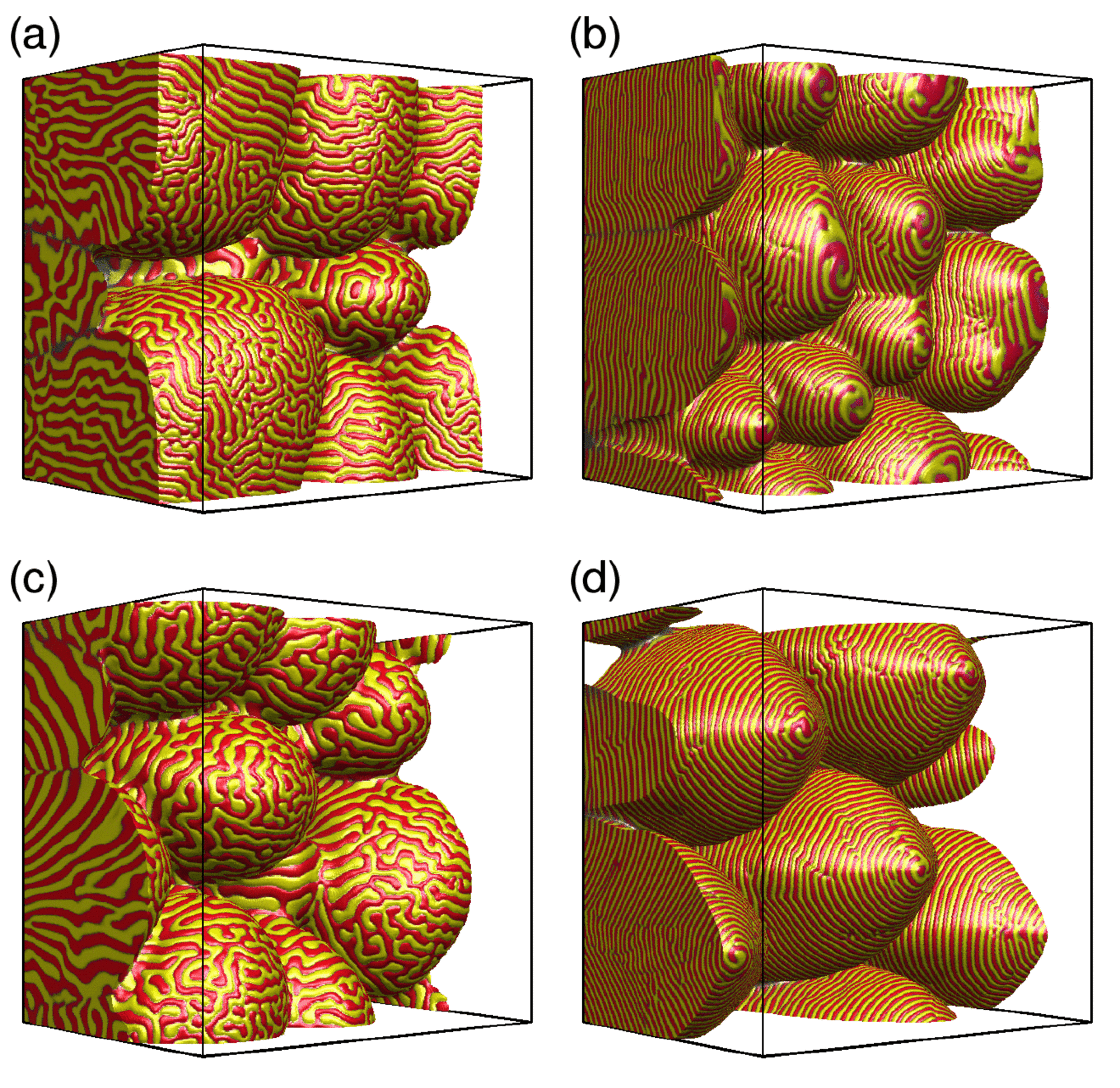}
  \caption{\label{fig:ani_vs_noani} The effect of velocity and anisotropy on the eutectic morphology grown from 25 single-phase seeds in large scale simulations ($380 \times 380 \times 511$ grid). (a), (b): without anisotropy, and (c), (d): with kinetic anisotropy of $\epsilon_4 = 0.3$. The pulling velocity is $v_p = 0.1$ on the left, and $v_p = 0.2$ on the right. Isotropic cases: In panels (a) and (b) a "random eutectic cellular" structure emerged that continuously changes with time, however, with lamellae perpendicular to the interface for the lower velocity (a), and lamellae nearly perpendicular to the temperature gradient at the higher velocity (b), yielding contourline-like patterns.  Anisotropic cases: In panel (c) a cellular structure forms, whereas in (d) a structure composed of spiraling eutectic dendrites evolves, which tends towards a final configuration that is considerably more stable than the cellular structure. [Panels (a)--(c) display snapshots taken at $3.4 \times 10^6, 3.2 \times 10^6, 1.66 \times 10^6,$ and $3.0 \times 10^6$ time steps, respectively.]}
\end{figure*}

\subsection{Is anisotropy needed for eutectic dendrites?}
One of the most interesting implications of the work of Akamatsu {\it et al.} \cite{akamatsu_2014} is that unlike in the case of single-phase dendrites, where the presence of anisotropy is a precondition of dendritic growth \cite{caroli_1986,benamar_1986,barbieri_1987,kessler_1988,saito_1988,brener_1992,ihle_1993,ihle_1994}, it is expected that in the case of spiraling ternary eutectic dendrites dendritic growth is possible without anisotropy. In order to clarify this issue further, we have performed simulations for $\epsilon_4 = 0$. To minimize finite size effects that, in principle, might contribute to the realization of the two-phase dendrites, we have made large cross-section  simulations (on $512 \times 512 \times 200$, $380 \times 380 \times 511$, and $256 \times 256 \times 800$ grids) at the CCMS, IMR, Tohoku University. These simulations were started from 25 to 100 randomly positioned single-phase seeds. Performing the simulation without anisotropy at $v_p = 0.1$, the growing eutectic particles impinge upon each other, and after a transient period of frustrated eutectic patterns, the solid-liquid interface forms a 3D disordered cellular morphology with dynamically appearing and disappearing tips and ridges of rather flat end, covered by a disordered lamellar eutectic pattern with lamellae perpendicular to the solid-liquid interface (Fig.~\ref{fig:large}); patterns well known as eutectic colonies seen in ternary systems experimentally \cite{akamatsu_2000} and theoretically \cite{plapp_1999, plapp_2002}. The effect of pulling velocity and anisotropy are illustrated in Fig.~\ref{fig:ani_vs_noani}. The growth forms in the isotropic systems are shown in Figs.~\ref{fig:ani_vs_noani}(a) and (b). In both cases dynamically changing cellular structures are observed, however, for the larger velocity the "lamellae" are close to perpendicular to the temperature gradient, indicating a nearly 1D eutectic growth expected at extreme high velocities \cite{lu_2012}; a mechanism possibly emerging from the fact that at large undercoolings the applied model has a spinodal mechanism to form the solid phases upon each other. While in case (b) {\it local spiraling may appear temporarily at the tips}, well developed steady-state spiraling dendrites cannot be seen. In these isotropic cases, the larger flat tips tend to undergo a tip-splitting/tip elimination phenomenon akin to the dynamic tip splitting and cell elimination process shown by quantitative phase-field modeling of cellular growth in binary alloys of low anisotropy \cite{gurevich_2010, ma_2014}. The respective simulations performed with kinetic anisotropy of $\epsilon_4 = 0.3$ are displayed in Figs.~\ref{fig:ani_vs_noani} (c) and (d). While for the smaller velocity a dynamically changing cellular structure is observed, at the higher velocity [Fig.~\ref{fig:ani_vs_noani}(d)], the peaks are far more regular than in the other cases and converge to a stable dendritic configuration. Remarkably, we do not see side branches even in these large simulations. (We suspect that considerably larger simulation box is needed to capture this feature.) It appears that the spiraling eutectic pattern is considerably more definite with anisotropy than without anisotropy [{\it cf.} Figs.~\ref{fig:ani_vs_noani}(a), (b), and (c) to Fig.~\ref{fig:ani_vs_noani}(d)]. Note that these results, do not rule out the possibility that in the absence of anisotropy spiraling steady-state eutectic dendrites occur as excessively rare events. Large scale computations are needed to clarify this issue further.

\section{Summary}
We have investigated the formation of ternary eutectic dendrites within the phase-field theory. It has been shown that a minimal ternary phase-field model of eutectic solidification suffices for reproducing the spiraling eutectic dendrites. We have mapped the domain of the Gibbs simplex, in which eutectic dendrites of ordered patterns (spiraling or target) appear. Increasing the pulling velocity, we see the following sequence of transitions between morphologies/patterns: flat front lamellae $\rightarrow$ eutectic colonies $\rightarrow$ eutectic dendrites $\rightarrow$ dendrites with target pattern $\rightarrow$ partitionless dendrites (due to solute trapping for components A and B, but not for C) $\rightarrow$ flat partitionless solid (due to full solute trapping). We have confirmed the assumptions of Akamatsu {\it et al.} \cite{akamatsu_2014} that (i) the Jackson-Hunt scaling is followed by the spiraling eutectic pattern; and (ii) that the eutectic wavelength and the tip radius of the dendrite are proportional and of comparable magnitude. The geometrical shape of the eutectic dendrites and the tip radius behave analogously to their single-phase counterparts. Apparently, the underlying eutectic pattern has little influence on the shape of the dendrite.  A number of quasi steady-state eutectic patterns have been observed under nominally the same conditions, including the target pattern, and single to multiple spirals, of which the fluctuations choose. This stochastic behavior resembles to that of helical Liesegang systems \cite{thomas_2013b}, raising the possibility of a universality class for systems displaying spiraling under diffusion control.  It remains to be seen, however, whether this stochastic behavior is indeed general. In spiraling dendrites the $\alpha$ (and also the $\beta$) phase domains are continuous throughout the solidified regions. It appears that the number of the spiral arms increases with increasing tip radius. The expectation that no anisotropy is needed for the formation of two-phase dendrites \cite{akamatsu_2014} has not been born out by our simulations: Without anisotropy, the emerging cellular structure does not display regular steady-state spiraling. It cannot be excluded, however, that in the isotropic systems spiraling eutectic dendrites are rare events that cannot be easily captured by simulations.

\acknowledgments{This work has been supported by the EU FP7 projects ``ENSEMBLE" (Grant Agreement NMP4-SL-2008-213669) and ``EXOMET" (contract No. NMP-LA-2012-280421, co-founded by ESA). We thank Mathis Plapp and Zolt\'an R\'acz for the enlightening discussions. The authors gratefully acknowledge the use of the SR16000 supercomputing facility at the Center for Computational Materials Science of the Institute for Materials Research, Tohoku University, Sendai, Japan.}


\begin{thebibliography}{99}
\bibitem{lambconst} This finding ($\lambda \approx$ const.) follows from the fact that in the present model the free energy of the solid-solid interface is independent of the solid-liquid interface free energy. This and the symmetric features of the model system adopted here leads to a situation in which the effect of the variation of the solid-liquid interface free energy on the Jackson-Hunt wavelength is compensated by a simultaneous change of the contact angles.
\end{thebibliography}
\end{document}